\documentclass[a4paper,11pt]{article}
\usepackage{graphicx}
\usepackage{comment}
\usepackage{amssymb}
\usepackage{amsmath}
\usepackage{nicefrac}
\usepackage{graphicx}
\usepackage{dcolumn}
\usepackage{bm}
\usepackage{comment}
\begin{document}

\newcommand{\bin}[2]{\left(\begin{array}{c} \!\!#1\!\! \\  \!\!#2\!\! \end{array}\right)}
\newcommand{\troisj}[3]{\left(\begin{array}{ccc}#1 & #2 & #3 \\ 0 & 0 & 0 \end{array}\right)}
\newcommand{\troisjm}[6]{\left(\begin{array}{ccc}#1 & #2 & #3 \\ #4 & #5 & #6 \end{array}\right)}
\newcommand{\sixj}[6]{\left\{\begin{array}{ccc}#1 & #2 & #3 \\ #4 & #5 & #6 \end{array}\right\}}
\newcommand{\neufj}[9]{\left\{\begin{array}{ccc}#1 & #2 & #3 \\ #4 & #5 & #6 \\ #7 & #8 & #9 \end{array}\right\}}
\newcommand{\eline}{E_{\gamma J,\gamma'J'}}

\huge

\begin{center}
Characterization of anomalous Zeeman patterns in complex atomic spectra
\end{center}

\vspace{0.5cm}

\large

\begin{center}
Jean-Christophe Pain\footnote{jean-christophe.pain@cea.fr (corresponding author)} and Franck Gilleron
\end{center}

\vspace{0.2cm}

\normalsize

\begin{center}
CEA, DAM, DIF, F-91297 Arpajon, France
\end{center}

\vspace{0.5cm}

\begin{abstract}
The modeling of complex atomic spectra is a difficult task, due to the huge number of levels and lines involved. In the presence of a magnetic field, the computation becomes even more difficult. The anomalous Zeeman pattern is a superposition of many absorption or emission profiles with different Zeeman relative strengths, shifts, widths, asymmetries and sharpnesses. We propose a statistical approach to study the effect of a magnetic field on the broadening of spectral lines and transition arrays in atomic spectra. In this model, the $\sigma$ and $\pi$ profiles are described using the moments of the Zeeman components, which depend on quantum numbers and Land\'{e} factors. A graphical calculation of these moments, together with a statistical modeling of Zeeman profiles as expansions in terms of Hermite polynomials are presented. It is shown that the procedure is more efficient, in terms of convergence and validity range, than the Taylor-series expansion in powers of the magnetic field which was suggested in the past. Finally, a simple approximate method to estimate the contribution of a magnetic field to the width of transition arrays is proposed. It relies on our recently published recursive technique for the numbering of LS-terms of an arbitrary configuration.
\end{abstract}

\section{\label{sec1} Introduction}

In astrophysics, the observation of a splitting of spectral lines in the visible and UV ranges for a few white dwarfs \cite{BABCOCK60} confirmed the existence of intense magnetic fields (0.1 - 10$^4$ MG) as predicted by Blackett \cite{BLACKETT47}. The influence of a magnetic field on an atom modifies its emission or absorption lines. Thanks to this property, known as Zeeman effect, the detection of magnetic fields is possible at large distances, through the measured radiation. The linear and quadratic Zeeman effects \cite{ZEEMAN1896a,ZEEMAN1897} explain the separation of spectral lines and enable one to determine a value of the magnetic field. In the same way, pulsars and neutron stars having an even more intense magnetic field (10$^5$ - 10$^8$ MG) have been discovered through their spectrum in the range of radio-frequencies and X-rays. There are numerous astrophysical applications, either direct or indirect, and requiring sometimes a sophisticated theoretical modeling. The methods differ according to the nature of the objects studied (see table \ref{tab1}), the magnitude and the geometry of the magnetic fields, and to the quality of the observation in terms of sensitivity and spectral resolution. Moreover, the variations of the magnetic field of stars during their rotation bring some information about their global geometry. The ``spectro-polarimetric'' methods exploit the additional recording of the circular polarization with respect to the wavelength. This enables one to obtain a detailed map of the field \cite{RYABCHICOVA06} through a separation of its components parallel or perpendicular to the line of sight.

Strong magnetic fields are also encountered, for instance, in magneto-inertial fusion \cite{LINDMAN10}. Inserting a magnetic field into inertial-confinement-fusion capsules before compressing them \cite{KIRKPATRICK95} presents the advantages to suppress the electron thermal-conduction losses and to better control the $\alpha$-particle energy deposition. The magnetic fields generated inside a Hohlraum can reach a few MG.

\begin{table}[ht]
\begin{center}
\begin{tabular}{|c|c|} \hline\hline
Magnetic field $B$ (MG) & Astrophysical object \\ \hline\hline
10$^5$ - 10$^8$ & Neutron star or pulsar \\ 
10$^{-1}$ - 10$^4$ & White dwarf \\ 
10$^{-4}$ - 10$^{-2}$ & Hot magnetic star \\ 
0 - 10$^{-6}$ & Planets of the solar system \\ 
10$^{-13}$ - 10$^{-11}$ & Interstellar cloud \\ \hline\hline
\end{tabular}
\end{center}
\caption{Orders of magnitude of magnetic fields encountered in astrophysics (1 MG=10$^6$ G=100 T).}\label{tab1}
\end{table}

In this work, the effect of a magnetic field on the broadening of spectral lines and transition arrays in complex atomic spectra is investigated. A proper description of physical broadening mechanisms \cite{STAMBULCHIK10} requires a simultaneous treatment of Stark and Zeeman effects, which was performed by Ferri \emph{et al.} \cite{FERRI11} in the framework of the Frequency Fluctuation Model \cite{CALISTI10}. In the case of an atom (ion) having several open sub-shells, the number of electric dipolar lines can be immense and the anomalous Zeeman pattern is a superposition of many profiles. When dealing with a huge number of simultaneously recorded profiles, it becomes necessary to characterize the line shape in terms of a limited number of parameters, and therefore to determine constraints on modelings. A statistical analysis can be performed using the moments of the profile. The $n^{th}$-order centered moment $\mu_{n,c}[A]$ of a distribution $A(E)$ is defined by 

\begin{equation}
\mu_{n,c}[A]=\frac{\int_{-\infty}^{\infty}(E-\mu_1)^nA(E)~dE}{\int_{-\infty}^{\infty}A(E)~dE},
\end{equation}

where

\begin{equation}
\mu_1=\frac{\int_{-\infty}^{\infty}E~A(E)~dE}{\int_{-\infty}^{\infty}A(E)~dE}
\end{equation}

is the center of gravity of $A(E)$. Each absorption or emission profile constituting the anomalous Zeeman pattern has its own strength, shift (first-order moment), width (second-order moment), asymmetry (third-order moment) and sharpness (fourth-order moment). We discuss different ways of calculating these moments (whatever the order) in terms of the quantum numbers and Land\'{e} factors of the levels involved in the line and present a statistical modeling of the Zeeman profile. It relies on the use of a A-type Gram-Charlier expansion series for each of the components $\Delta M$=0, $+1$ and $-1$. Finally, leaning on our recently published recursive approach for the numbering of LS-terms of an arbitrary configuration \cite{GILLERON09}, we propose a simple approximation to estimate the contribution of a magnetic field to the emission and absorption coefficients.

The paper is organized as follows. In section \ref{sec2}, the intensity distribution of an electric-dipolar (E1) line is introduced, together with its strength-weighted moments. In section \ref{sec3}, a graphical representation of the angular-momentum sum rules involved in the calculations of the moments is described. It reveals the way the Racah algebra proceeds and is simple to compute: the $n^{th}$-order moment reduces to a regular polygon with $(n+2)$ sides. In section \ref{sec4}, the statistical modeling of a line perturbed by a magnetic field is discussed, using particular distributions involving the reduced centered moments of the Zeeman $\pi$ and $\sigma_{\pm}$ components. It is proven that the Gram-Charlier development is more efficient than the usual Taylor-series expansion. In section \ref{sec5}, an efficient approach to take into account the effect of a magnetic field on a transition array is proposed. In section \ref{sec6} it is shown that the techniques presented in this paper still apply when hyperfine interaction is included and section \ref{sec7} is the conclusion. 

\section{\label{sec2} Intensities and characteristics of Zeeman components}

The Zeeman Hamiltonian reads:

\begin{equation}
H_Z=\mu_BB\;(L_z+g_sS_z),
\end{equation}

where $B$ is the magnitude of the magnetic field along the $z$-axis $\vec{B}=B\;\vec{u}_z$, $\mu_B$ the Bohr magneton, $g_s=2.0023192$ is the anomalous gyromagnetic ratio for the electron spin, and $L_z$ and $S_z$ respectively the projections of total orbital and spin angular momenta of the system. For sufficiently weak values of the field $B$, the off-diagonal matrix elements of $H_Z$ that connect basis states of different values of $J$ (modulus of the total angular-momentum of the system $\vec{J}=\vec{L}+\vec{S}$) will be negligible compared to the contributions of the Coulomb and spin-orbit interactions to the energy. It becomes then reasonable to neglect the mixing of basis states of different values of $J$. The energy matrix breaks down into blocks according to the value of $J$ (as in the field-free case) and the contribution of the magnetic field to the energy can be calculated as a simple perturbation. The following expression for the diagonal matrix element of $H_Z$ for the state $|\gamma JM\rangle$

\begin{equation}
\langle\gamma J M|L_z+g_sS_z|\gamma J M\rangle=g_{\gamma J}\langle\gamma J M|J_z|\gamma J M\rangle=g_{\gamma J} M,
\end{equation}

where $J_z=L_z+S_z$, defines the Land\'e factor $g_{\gamma J}$ of level $\gamma J$ \cite{LANDE21}. One can roughly consider that Zeeman approach is no longer valid when the magnetic field is of the same order of magnitude as the spin-orbit contribution (see table \ref{tab2}):

\begin{equation}
B_c=(Z^*e^2/\hbar c)^2me^4/(\mu_B\hbar^2).
\end{equation}

In that case, a Paschen-Back \cite{PASCHEN21} treatment is necessary.

\begin{table}[ht]
\begin{center}
\begin{tabular}{|c|c|}\hline\hline
Element & $B_c(MG)$ \\\hline\hline
H (Z=1) & 0.0078 \\ 
Al (Z=13) & 1.30 \\ 
Ni (Z=28) & 6.10 \\ 
Nb (Z=41) & 13.10 \\ 
Sm (Z=62) & 30.00 \\ 
Po (Z=84) & 55.00 \\
Np (Z=93) & 67.50 \\\hline\hline
\end{tabular}
\end{center}
\caption{Critical value of the magnetic field for which the spin-orbit interaction becomes of the same order of magnitude as the magnetic interaction. This gives an estimate of the critical field for which Paschen-Back treatment is more appropriate than Zeeman description.}\label{tab2}
\end{table}

In the presence of a magnetic field, the total intensity of transition $\gamma JM\rightarrow\gamma'J'M'$ at the energy $E$ reads:

\begin{eqnarray}
I(E)&=&\sum_{\gamma JM\rightarrow\gamma'J'M'}S_{\gamma JM,\gamma'J'M'}\nonumber\\
& &\times\Psi_{\gamma JM,\gamma'J'M'}(E-E_{\gamma JM,\gamma'J'M'}),
\end{eqnarray}

where

\begin{equation}
E_{\gamma JM,\gamma'J'M'}=\eline+\mu_BB~(g_{\gamma' J'}M'-g_{\gamma J}M)
\end{equation}

and $S_{\gamma JM,\gamma'J'M'}$ are respectively the energy and the strength of a transition $\gamma JM\rightarrow\gamma'J'M'$. $\eline$ represents the energy of the line $\gamma J\rightarrow\gamma'J'$:

\begin{equation}
\eline=E_{\gamma'J'}-E_{\gamma J}=\langle\gamma'J'|H|\gamma'J'\rangle-\langle\gamma J|H|\gamma J\rangle, 
\end{equation}

where $H$ is the Hamiltonian of the system. The normalized profile $\Psi_{\gamma JM,\gamma'J'M'}(E)$ takes into account the broadening of the line due to radiative decay, Doppler effect, ionic Stark effect, electron collisions, \textit{etc.}

Assuming that the optical media is passive (\emph{e.g.} there is no Faraday rotation), the intensity, detected with an angle of observation $\theta$, is given by \cite{GODBERT09,NGUYENHOE67}:

\begin{equation}\label{itot}
I_{\theta}(E)=I_{\parallel}(E)\cos^2(\theta)+I_{\bot}(E)\sin^2(\theta),
\end{equation}

where the longitudinal intensity is

\begin{equation}
I_{\parallel}(E)=\frac{1}{2}\left(I_{+1}(E)+I_{-1}(E)\right)
\end{equation}

and the transverse intensity

\begin{equation}
I_{\bot}(E)=\frac{1}{4}\left(I_{+1}(E)+I_{-1}(E)+2I_0(E)\right).
\end{equation}

$I_{\theta}(E)$ can be written in the form

\begin{eqnarray}
I_{\theta}(E)&=&\left(\frac{1+\cos^2(\theta)}{4}\right)\left(I_{+1}(E)+I_{-1}(E)\right)\nonumber\\
& &+\frac{\sin^2(\theta)}{2}I_0(E).
\end{eqnarray}

Each line $\gamma J\rightarrow\gamma'J'$ can be represented as a sum of three helical components associated to the selection rules $M'$=$M+q$, where the polarization $q$ is equal to 0 for $\pi$ components and to $\pm 1$ for $\sigma_{\pm}$ components. The intensity of the $q$ component of the E1 line $\gamma J\rightarrow\gamma'J'$ reads, assuming that all quantum states are populated in the statistical-weight approximation (high-temperature limit):

\begin{eqnarray}\label{defiq}
I_q(E)&=&\sum_{M,M'}S_{M,M',q}\nonumber\\
& &\times\Psi_{\gamma JM,\gamma'J'M'}(E-E_{\gamma JM,\gamma'J'M'}),
\end{eqnarray}

where

\begin{equation}
S_{M,M',q}=C_{M,M',q}\times S_{\gamma J,\gamma'J'}
\end{equation}

and

\begin{equation}
C_{M,M',q}=3\troisjm{J}{1}{J'}{-M}{-q}{M'}^2.
\end{equation}

The quantity

\begin{equation}
S_{\gamma J,\gamma'J'}=\sum_{\gamma JM\rightarrow\gamma'J'M'}S_{\gamma JM,\gamma'J'M'}
\end{equation}

represents the strength of the line $\gamma J\rightarrow\gamma'J'$ and is proportional to $|\langle \gamma J|\mathcal{Z}|\gamma'J'\rangle|^2$, where $\mathcal{Z}$ is the $z$ component of the dipole transition operator. Since

\begin{equation}
\sum_{M,M'}\troisjm{J}{1}{J'}{-M}{-q}{M'}^2=\frac{1}{3},
\end{equation}

each component has the same strength. The number of transitions in each component is equal to 2$\times\min(J,J')$+1. The distribution $I_q(E)$ can be characterized by the moments centered in $\eline$:

\begin{eqnarray}
\mathcal{M}_k^{[q]}&=&\sum_{M,M'}C_{M,M',q}~(g_{\gamma'J'}M'-g_{\gamma J}M)^k\nonumber\\
&=&\sum_{M,M'}C_{M,M',q}\nonumber\\
& &\times\left(g_{\gamma'J'}M'-g_{\gamma J}M-\mathcal{M}_1^{[q]}+\mathcal{M}_1^{[q]}\right)^k\nonumber\\
&=&\sum_{i=0}^k\bin{k}{i}\mathcal{M}_{i,c}^{[q]}\;\left(\mathcal{M}_1^{[q]}\right)^{k-i},
\end{eqnarray}

where

\begin{equation}
\mathcal{M}_{n,c}^{[q]}=\sum_{M,M'}C_{M,M',q}~\left(g_{\gamma'J'}M'-g_{\gamma J}M-\mathcal{M}_1^{[q]}\right)^n
\end{equation}

is the $n^{th}$-order centered moment of the distribution. It is useful to introduce the reduced centered moments defined by

\begin{equation}
\alpha_n^{[q]}=\sum_{M,M'}C_{M,M',q}~\left(\frac{g_{\gamma'J'}M'-g_{\gamma J}M-\mathcal{M}_1^{[q]}}{\sqrt{\mathcal{V}^{[q]}}}\right)^n,
\end{equation}

where $\mathcal{M}_1^{[q]}$ is the center-of-gravity of the strength-weighted component energies (relative to $\eline$ and in units of $\mu_BB$) and $\sqrt{\mathcal{V}^{[q]}}=\sqrt{\mathcal{M}_{2,c}^{[q]}}$ is the standard deviation (in units of $\mu_BB$). Centered moments of $\sigma_-$ and $\sigma_+$ components are related by $\mathcal{M}_{n,c}^{[\sigma_-]}=(-1)^n\mathcal{M}_{n,c}^{[\sigma_+]}$. The use of $\alpha_n^{[q]}$ instead of $\mathcal{M}_n^{[q]}$ allows one to avoid numerical problems due to the occurence of large numbers. The first values are $\alpha_0^{[q]}=1$, $\alpha_1^{[q]}=0$ and $\alpha_2^{[q]}=1$. The distribution $I_q(E)$ is therefore fully characterized by the values of $\mathcal{M}_1^{[q]}$, $\mathcal{V}^{[q]}$ and of the high-order moments $\alpha_n^{[q]}$ with $n>2$. It is reasonable to consider that the first four moments are sufficient to capture the global shape of the distribution $I_q(E)$ (see for instance Ref. \cite{KENDALL69}, p.88-89). The third- and fourth-order reduced centered moments $\alpha_3^{[q]}$ and $\alpha_4^{[q]}$ are named {\it skewness} and {\it kurtosis}. They quantify respectively the asymmetry and sharpness of the distribution. The kurtosis is usually compared to the value $\alpha_4^{[q]}=3$ for a Gaussian.

\section{\label{sec3} Moments of the Zeeman components $\pi$, $\sigma_+$ and $\sigma_-$ of a line $\gamma J\rightarrow\gamma'J'$}

\subsection{\label{subsec31} Racah algebra and graphical representation}

The moments can be easily derived using Racah algebra and graphical techniques \cite{VARSHALOVICH88,JUCYS62,ELBAZ69,ELBAZ72,ELBAZ85}. We define the notations $[a, b, c, \cdots]=(2a+1)(2b+1)(2c+1)\cdots$, and use the convention of Biedenharn {\it et al.}: $\bar{x}=x(x+1)$ \cite{BIEDENHARN52}. Since $M$ can be expressed as

\begin{equation}
M=(-1)^{J-M}\sqrt{[J]\bar{J}}\troisjm{J}{1}{J}{-M}{0}{M},
\end{equation}

the first-order moment can be obtained from the relations (\ref{rs1}), (\ref{tj1}) and (\ref{sj2}) given in appendix A \cite{VARSHALOVICH88}. One has

\begin{equation}
\sum_{M,M'}\troisjm{J}{1}{J'}{-M}{-q}{M'}^2M=-\frac{q}{12}\left(\bar{J}-\bar{J'}+2\right),
\end{equation}

and

\begin{equation}
\sum_{M,M'}\troisjm{J}{1}{J'}{-M}{-q}{M'}^2M'=\frac{q}{12}\left(\bar{J'}-\bar{J}+2\right),
\end{equation}

which gives finally \cite{SHENSTONE29}

\begin{eqnarray}
\mathcal{M}_1^{[q]}&=&\frac{q}{4}\;\left[2(g_{\gamma J}+g_{\gamma'J'})\right.\nonumber\\
& &\left.+(g_{\gamma J}-g_{\gamma'J'})(J-J')(J+J'+1)\right].
\end{eqnarray}

\begin{figure}
\begin{center}
\reflectbox{\includegraphics[width=2.6cm]{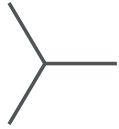}}
\end{center}
\caption{Graphical representation of a three-$jm$ coefficient.}
\label{fig1}
\end{figure}

\begin{figure}
\begin{center}
\reflectbox{\includegraphics[width=2.3cm]{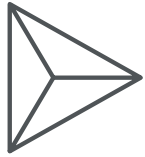}}
\end{center}
\caption{Graphical representation of a six-$j$ coefficient.}
\label{fig2}
\end{figure}

\begin{figure}
\begin{center}
\includegraphics[width=10.6cm]{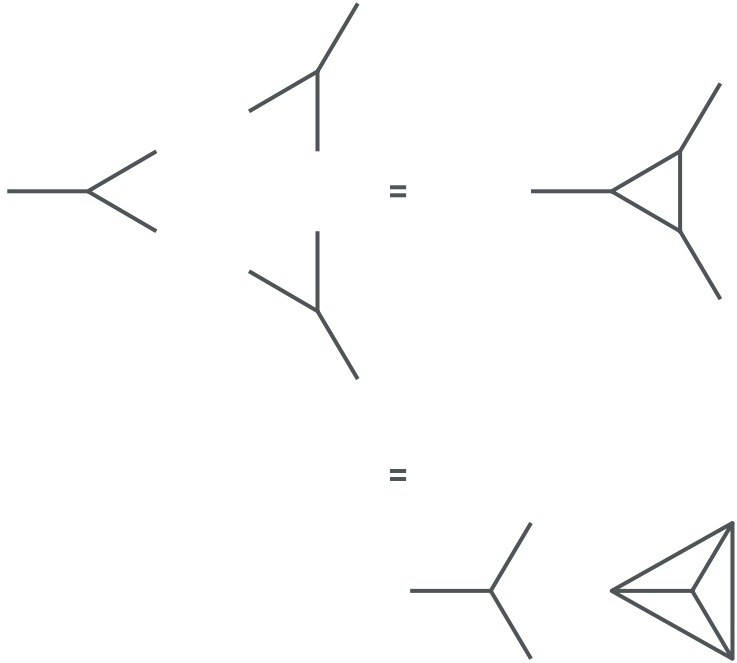}
\end{center}
\caption{Simplified graphical representation of the sum rule (\ref{rs1}) of appendix A involved in the calculation of the first-order moment $\mathcal{M}_1^{[q]}$ of a Zeeman component. The first equality corresponds to the merging of the three three-$jm$ symbols, and the second one the splitting into a three-$jm$ and a six-$j$ symbol.}
\label{fig3}
\end{figure}

\begin{figure}
\begin{center}
\reflectbox{\includegraphics[width=10.6cm]{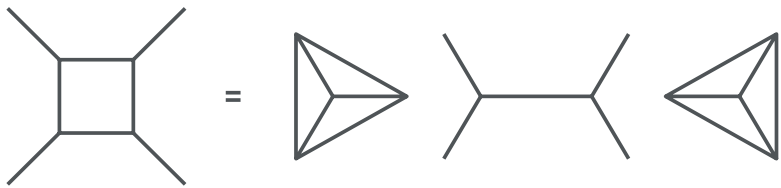}}
\end{center}
\caption{Simplified graphical representation of the sum rule (\ref{rs2}) of appendix A involved in the calculation of the second-order moment $\mathcal{M}_2^{[q]}$ of a Zeeman component.}
\label{fig4}
\end{figure}

\begin{figure}
\begin{center}
\reflectbox{\includegraphics[width=10.6cm]{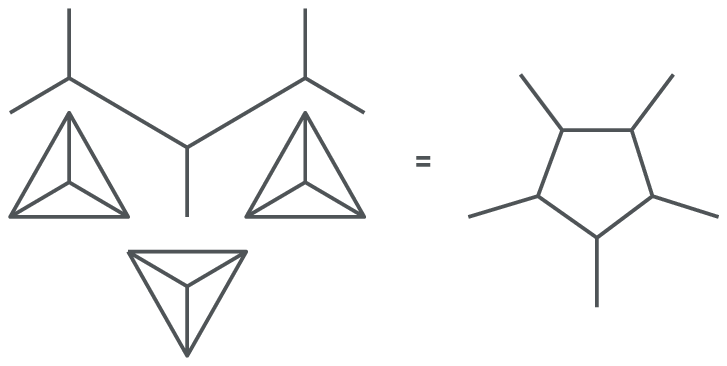}}
\end{center}
\caption{Simplified graphical representation of the sum rule involved in the calculation of the third-order moment $\mathcal{M}_3^{[q]}$ of a Zeeman component.}
\label{fig5}
\end{figure}

\begin{figure}
\begin{center}
\reflectbox{\includegraphics[width=10.6cm]{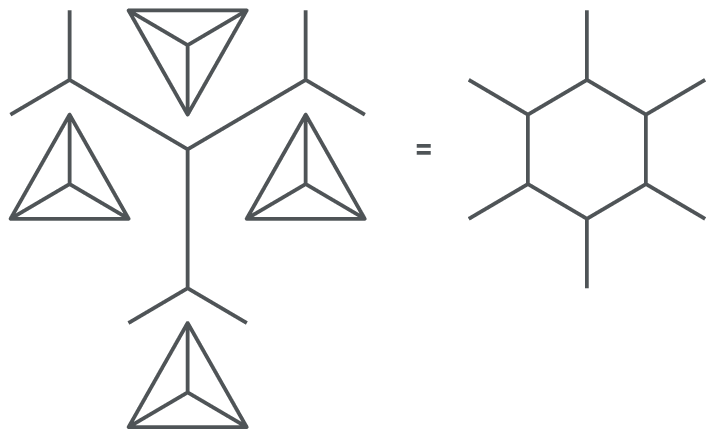}}
\end{center}
\caption{Simplified graphical representation of the sum rule involved in the calculation of the fourth-order moment $\mathcal{M}_4^{[q]}$ of a Zeeman component.}
\label{fig6}
\end{figure}

The variance is obtained using the sum rule (\ref{rs2}) \cite{VARSHALOVICH88} together with the expressions (\ref{tj1}) to (\ref{tj5}) \cite{EDMONDS57,LANDI85,BAUCHE88b}. More generally, the $n^{th}$-order moment involves the following sum rule:

\begin{equation}
\sum_{M,M'}(-1)^{n(J-M)}\troisjm{J}{1}{J'}{-M}{-q}{M'}^2\troisjm{J}{1}{J}{-M}{0}{M}^n,
\end{equation}

where $n$ is an integer. Figures \ref{fig1} and \ref{fig2} give the graphical simplified representations of a three-$jm$ and a six-$j$ symbol respectively. Each line represents an angular momentum \cite{VARSHALOVICH88,JUCYS62,ELBAZ69,ELBAZ72,ELBAZ85}. The names of the angular momenta (or of their projections in the case of three-$jm$ coefficients) and the phase factors are omitted. Figures \ref{fig3}, \ref{fig4}, \ref{fig5} and \ref{fig6} display the graphical representations of the calculations of the first four moments $\mathcal{M}_1^{[q]}$ and $\mathcal{M}_2^{[q]}$, $\mathcal{M}_3^{[q]}$ and $\mathcal{M}_4^{[q]}$ respectively. One can also see on Fig. \ref{fig3} how the three-$jm$ symbols merge into a single closed diagram. These schemes are a representation of summation rules and reduction formulas. Although some computer programs exist (see for instance \cite{KARAZIJA91a,KARAZIJA91b,KARAZIJA95,BORDARIER70,BAR88,OREG90}), which are devoted to the reduction of graphs, it is easy to understand that the calculation becomes more and more cumbersome as the order of the moment increases. The $n^{th}$-order moment reduces graphically to a polygone with ($n$+2) sides.

\subsection{\label{subsec32} Expression in terms of Bernoulli polynomials}

\begin{figure}[ht]
\begin{center}
\vspace{1cm}
\includegraphics[width=10.6cm]{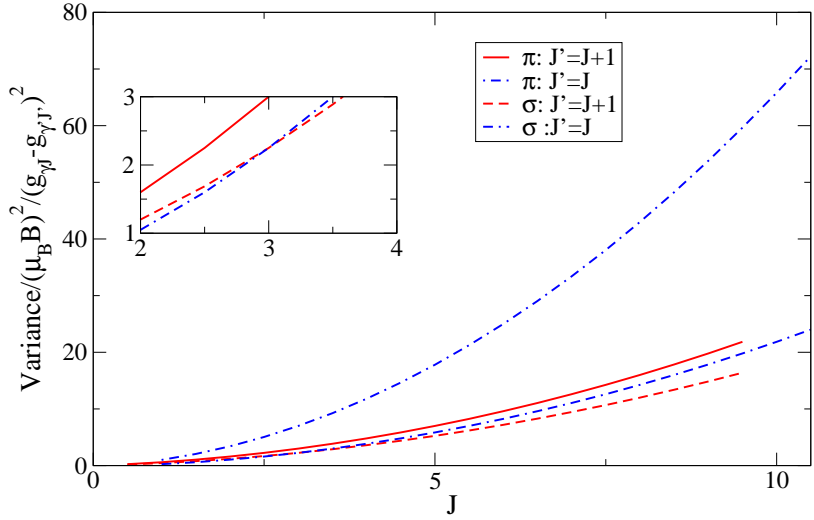}
\end{center}
\caption{(Color online) Variance of the $\pi$ and $\sigma$ components with respect to $J$.}
\label{fig7}
\end{figure}

\begin{figure}[ht]
\begin{center}
\vspace{1cm}
\includegraphics[width=10.6cm]{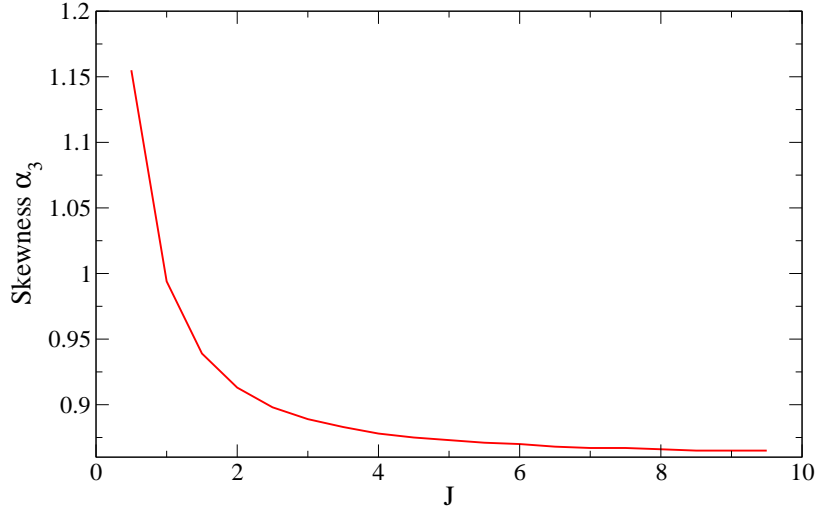}
\end{center}
\caption{Skewness $\alpha_3$ of the $\sigma_+$ component with respect to $J$ for $\Delta J=\pm 1$.}
\label{fig8}
\end{figure}

\begin{figure}[ht]
\begin{center}
\vspace{1cm}
\includegraphics[width=10.6cm]{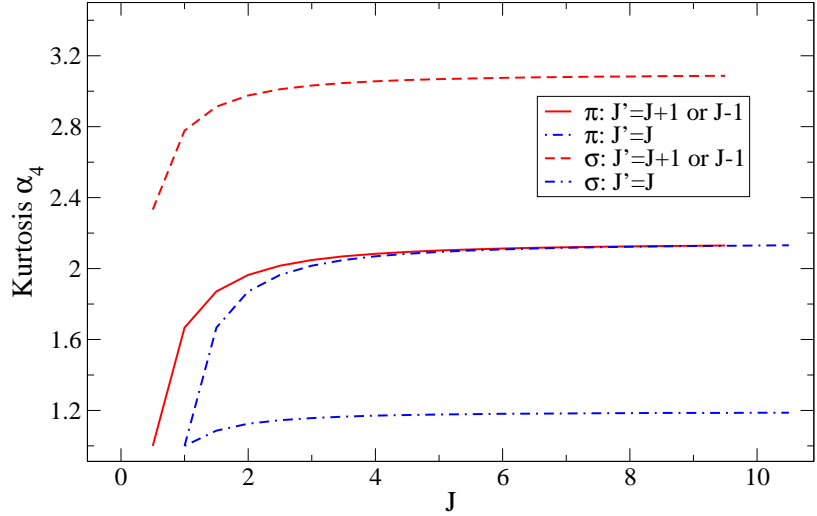}
\end{center}
\caption{(Color online) Kurtosis $\alpha_4$ of the $\pi$ and $\sigma$ components with respect to $J$ for $\Delta J$=0 and $\Delta J=\pm 1$.}
\label{fig9}
\end{figure}

\begin{figure}[ht]
\begin{center}
\vspace{1cm}
\includegraphics[width=10.6cm]{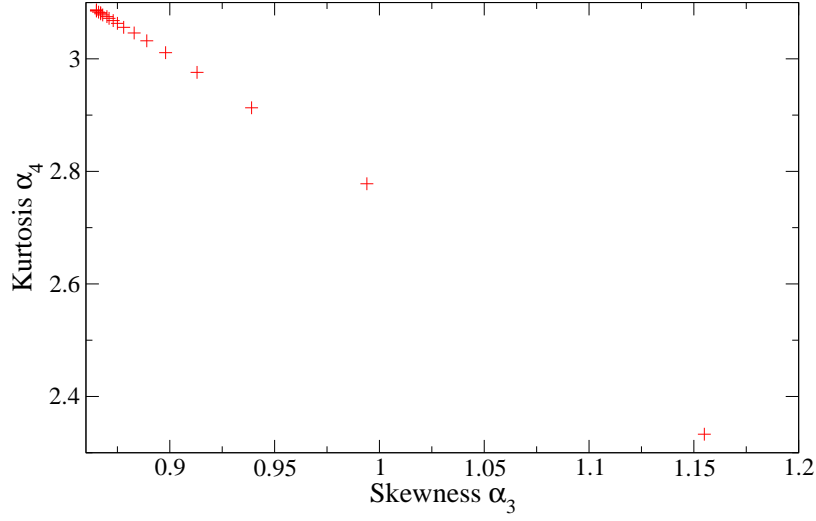}
\end{center}
\caption{Kurtosis $\alpha_4$ versus skewness $\alpha_3$ of the $\sigma_+$ component for $\Delta J$=+1.}
\label{fig10}
\end{figure}

\begin{figure}[ht]
\begin{center}
\vspace{1cm}
\includegraphics[width=10.6cm]{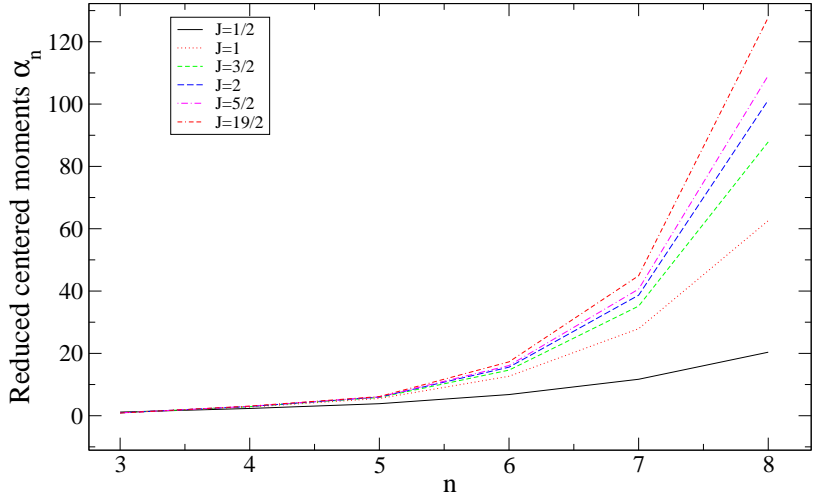}
\end{center}
\caption{(Color online) Reduced centered moments $\alpha_n$ of the $\sigma_+$ component versus $n$ for different values of $J$ in the case $\Delta J=\pm 1$.}
\label{fig11}
\end{figure}

Mathys and Stenflo \cite{MATHYS87a,MATHYS87b} have obtained more compact formulae for the moments in terms of Bernoulli polynomials (see appendix B). Values of $\alpha_3$ and $\alpha_4$ for the three selection rules $\Delta J=0,-1,+1$ are displayed in tables \ref{tab3} and \ref{tab4}. One finds that the variance of the $\pi$ component is always larger than the variance of the $\sigma_+$ and $\sigma_-$ components, indeed:

\begin{equation}\label{ineg1}
\mathcal{V}^{[\pi]}-\mathcal{V}^{[\sigma_{\pm}]}=\frac{(g_{\gamma J}-g_{\gamma'J'})^2}{20}(8\bar{J}-1)\;\;\mathrm{if}\;\;J'=J
\end{equation}

and

\begin{equation}\label{ineg2}
\mathcal{V}^{[\pi]}-\mathcal{V}^{[\sigma_{\pm}]}=\frac{(g_{\gamma J}-g_{\gamma'J'})^2}{20}(\bar{J}+J)\;\;\mathrm{if}\;\;J'=J+1,
\end{equation}

where $\mathcal{V}^{[q]}=\mathcal{M}_{2,c}^{[q]}=\mathcal{M}_{2}^{[q]}-\left(\mathcal{M}_{1}^{[q]}\right)^2$. Therefore, in all cases, $\mathcal{V}^{[\pi]}-\mathcal{V}^{[\sigma_{\pm}]}\geq$ 0. We can see on Fig. \ref{fig7} that the variance of the $\pi$ component for a given value of $J$ is larger for $\Delta J=0$ than for $\Delta J=\pm 1$ lines, and that the difference increases with $J$. Things are slightly different for the $\sigma$ components (see Fig. \ref{fig7}): the variance for $\Delta J=0$ overcomes the one from $\Delta J=\pm 1$ only for $J \ge$ 3. Moreover, the difference between both variances at fixed $J$ is smaller than for the $\pi$ component. Figure \ref{fig8} shows that the skewness $\alpha_3$ of the $\sigma_+$ component is a decreasing function of $J$ for $\Delta J=\pm 1$ line (the skewness is zero for $\Delta J=0$ since the splitting is symmetric in that case). On the contrary to the variance, the kurtosis $\alpha_4$ (see Fig. \ref{fig9}) is systematically higher for $\Delta J= \pm 1$ than for $\Delta J=0$, and the difference is almost constant and equal to 1. It is interesting to plot $\alpha_4$ versus $\alpha_3$ for the $\sigma_+$ component; it reveals that the dependence is quite linear, and that the values are very concentrated around 0.875 for the kurtosis and slightly above 3 for the skewness (see Fig. \ref{fig10}). As can be shown on Fig. \ref{fig11}, for a given value of $J$ the reduced centered moments $\alpha_n$ increase with the order $n$, and, for a given value of $n$, they increase as well with $J$, and get closer and closer when $J$ increases.

\begin{table}[ht]
\begin{center}
\begin{tabular}{|c|c|c|c|}\hline\hline 
$\sigma_+$ & $J'=J$ & $J'=J+1$ & $J'=J-1$ \\\hline\hline
$\alpha_3$ & 0 & $\frac{2\sqrt{5}}{3\sqrt{3}}\frac{J+1}{\sqrt{J(J+2)}}$ & $-\frac{2\sqrt{5}}{3\sqrt{3}}\frac{J}{\sqrt{J^2-1}}$ \\ 
$\alpha_4$ & $\frac{5}{7}\left(\frac{12\bar{J}-17}{4\bar{J}-3}\right)$ & $\frac{5}{21}\left(\frac{13J(J+2)-4}{J(J+2)}\right)$ & $-\frac{5}{21}\left(\frac{13J^2-17}{1-J^2}\right)$ \\\hline\hline
\end{tabular}
\end{center}
\caption{Values of $\alpha_3$ and $\alpha_4$ of the $\sigma_+$ component of E1 lines.}\label{tab3}
\end{table}

\begin{table}[ht]
\begin{center}
\begin{tabular}{|c|c|c|c|}\hline\hline 
$\pi$ & $J'=J$ & $J'=J+1$ & $J'=J-1$ \\\hline\hline
$\alpha_3$ & 0 & 0 & 0 \\ $\alpha_4$ & $\frac{25}{7}\left(\frac{3\{(J+2)J^2-1\}J+1}{\{1-3\bar{J}\}^2}\right)$ & $\frac{5}{7}\left(\frac{3J(J+2)-2}{J(J+2)}\right)$ & $\frac{5}{7}\left(\frac{3J^2-5}{J^2-1}\right)$ \\\hline\hline 
\end{tabular}
\end{center}
\caption{Values of $\alpha_3$ and $\alpha_4$ of the $\pi$ component of E1 lines.}\label{tab4}
\end{table}

The numerical values $\alpha_4$, $\alpha_6$ and $\alpha_8$ of the $\pi$ component for several lines are listed in table \ref{tab5}. Tables \ref{tab6} and \ref{tab7} contain the odd reduced centered moments of the $\sigma$ component for the same lines. 

\begin{table}[t]
\begin{center}
\begin{tabular}{|c|c|c|c|c|}\hline\hline
Line & $\mathcal{V}$ & $\alpha_4$ & $\alpha_6$ & $\alpha_8$\\ \hline\hline
$^5F_1\rightarrow~^5F_2$ & 0.60 & 1.667 & 2.778 & 4.629 \\ 
$^7D_3\rightarrow~ ^7D_4$ & 0.03 & 2.048 & 5.190 & 14.407 \\
$^4D_{3/2}\rightarrow~ ^4D_{5/2}$ & 1.05 & 1.871 & 3.944 & 8.436 \\ 
$^5P_2\rightarrow~ ^5P_3$ & 1.60 & 1.964 & 4.576 & 11.230 \\\hline\hline
\end{tabular}
\end{center}
\caption{Parameters of the $\pi$ component for several lines of the transition array Fe VII $3d^2\rightarrow 3d4p$. Even reduced centered moments.}\label{tab5}
\end{table}

\section{\label{sec4} Zeeman profile in low magnetic fields}

In the following, we consider the case where

\begin{equation}
I(E)=\frac{1}{3}\left(I_{+1}(E)+I_{-1}(E)+I_{0}(E)\right)
\end{equation}

which, according to Eq. (\ref{itot}), corresponds to an observation angle $\theta$ with $z$ axis such that $\cos^2(\theta)=\frac{1}{3}$.

\subsection{\label{subsec42} Taylor-series expansion}

In the following, we make the assumption that $\Psi_{\gamma J M,\gamma'J'M'}$ is a universal function $\Psi$ centered in $E_{\gamma J M,\gamma' J' M'}$. The quantity $I_q(E)$ (\ref{defiq}) can be expressed \cite{MATHYS87a,MATHYS87b} as a Taylor series around the line energy $E_{\gamma J,\gamma'J'}$:

\begin{eqnarray}
I_q(E)&=&S_{\gamma J,\gamma'J'}\times\left\{\vphantom{\frac{d^k}{dE^k}}\Psi(E-\eline)\right.\nonumber\\
& &+\sum_{k=1}^{\infty}\frac{(-1)^k}{k!}(\mu_BB)^k\mathcal{M}_k^{[q]}\nonumber\\
& &\left.\times\frac{d^k}{dE^k}\Psi(E-\eline)\right\}.
\end{eqnarray}

Assuming a Gaussian physical broadening of the lines:

\begin{equation}\label{gausbroa}
\Psi(E-\eline)=\frac{1}{\sqrt{2\pi v}}\exp\left(-\frac{(E-\eline)^2}{2v}\right),
\end{equation}

where $v$ represents the variance of the physical broadening mechanisms other than Zeeman effect (Doppler, Stark,...), we have (Rodrigues' formula):

\begin{eqnarray}
\frac{d^n}{dE^n}\Psi(E-\eline)&=&\frac{(-1)^n}{v^{n/2}}\Psi(E-\eline)\nonumber\\
& &\times\mathrm{He}_n\left(\frac{E-\eline}{\sqrt{v}}\right),
\end{eqnarray}

where $\text{He}_{k}$ is the Hermite polynomial of order $k$, related to the usual Hermite polynomial H$_k$ by

\begin{equation}
\mathrm{He}_k(x)=\frac{1}{2^{k/2}}\mathrm{H}_k\left(\frac{x}{\sqrt{2}}\right).
\end{equation}

He$_k$ obeys the recursion relation 

\begin{equation}
\mathrm{He}_{k+1}(x)=x~\mathrm{He}_{k}(x) - k~\mathrm{He}_{k-1}(x)
\end{equation}

with He$_0(x)$=1 and He$_1(x)=x$. The resulting expression of $I_q(E)$ reads

\begin{eqnarray}
I_q(E)&=&S_{\gamma J,\gamma'J'}\times\Psi(E-\eline)\nonumber\\
& &\times\left\{1+\sum_{k=1}^{\infty}\frac{(\mu_BB)^k}{k!\;v^{k/2}}~\mathcal{M}_k^{[q]}~\mathrm{He}_k\left(\frac{E-\eline}{\sqrt{v}}\right)\right\}.\nonumber\\
\end{eqnarray}

At the second order

\begin{eqnarray}\label{iq2}
I_q(E)&=&S_{\gamma J,\gamma'J'}\times\Psi(E-\eline)\nonumber\\
& &\times\left\{\vphantom{\frac{\left(v-\left(E-\eline\right)^2\right)}{2v^2}}1+\mu_BB\;\mathcal{M}_1^{[q]}\frac{(E-\eline)}{\sqrt{v}}\right.\nonumber\\
& &-(\mu_BB)^2\left(\mathcal{M}_{2,c}^{[q]}+\left(\mathcal{M}_1^{[q]}\right)\right)^2\nonumber\\
& &\left.\times\frac{\left(v-\left(E-\eline\right)^2\right)}{2v^2}\right\}.
\end{eqnarray}

\begin{figure}[ht]
\begin{center}
\vspace{1cm}
\includegraphics[width=10.6cm]{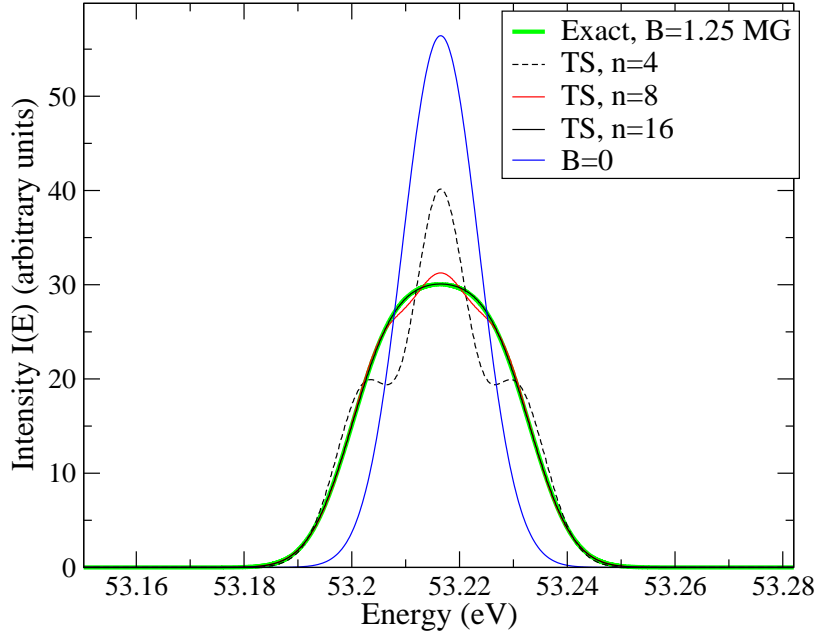}
\end{center}
\caption{(Color online) Modeling of a line $J=3\rightarrow\;J'=4$ of transition array Fe VII $3d^2\rightarrow 3d4p$ with Taylor-series expansion of different orders, compared to the exact calculation for B=1.25 MG and $v$=5 10$^{-5}$ eV$^2$. The development for $n$=16 and the exact calculation are superimposed.}
\label{fig12}
\end{figure}

\begin{figure}[ht]
\begin{center}
\vspace{1cm}
\includegraphics[width=10.6cm]{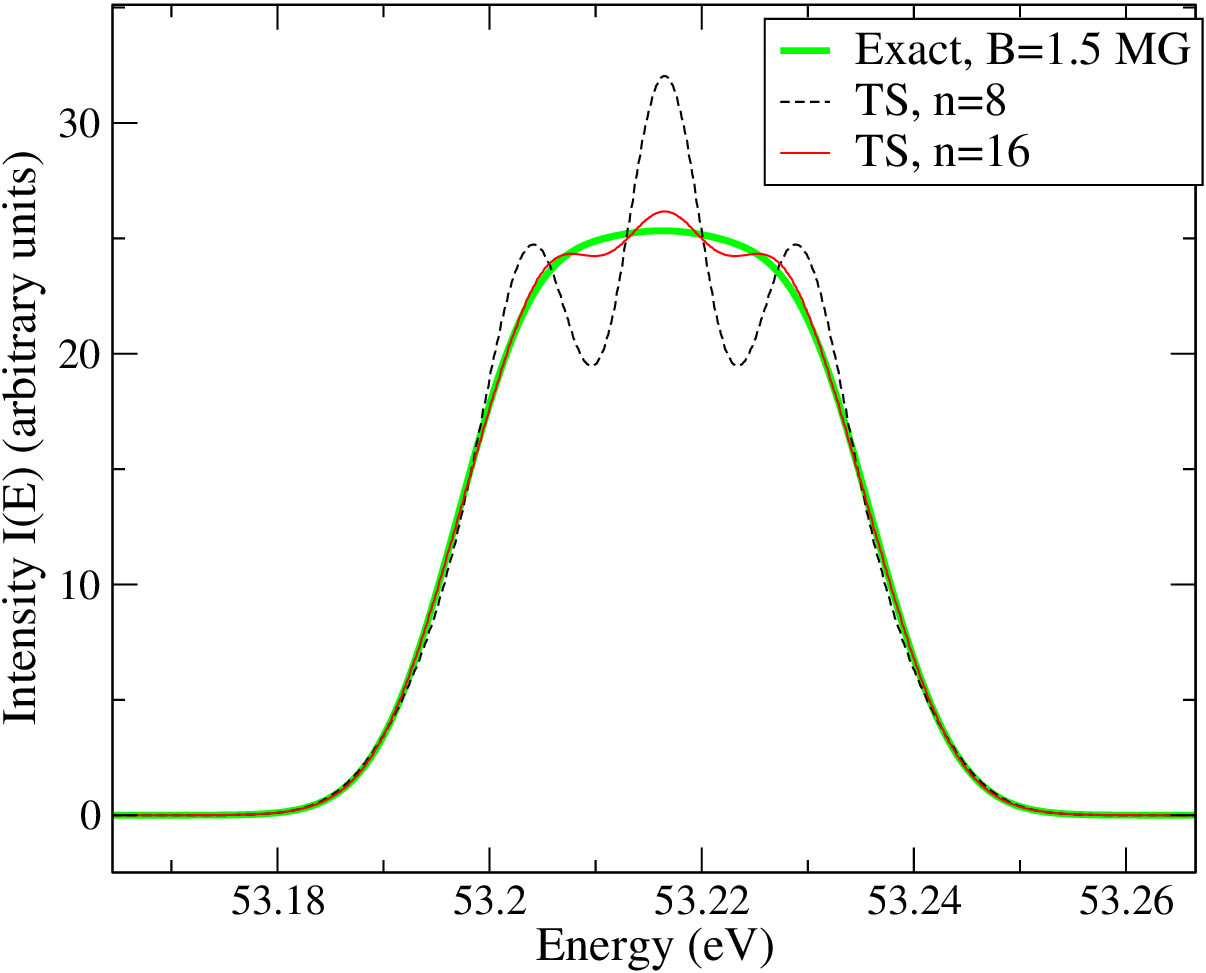}
\end{center}
\caption{(Color online) Modeling of a line $J=3\rightarrow\;J'=4$ of transition array Fe VII $3d^2\rightarrow 3d4p$ with Taylor-series expansion of different orders, compared to the exact calculation for B=1.5 MG and $v$=5 10$^{-5}$ eV$^2$.}
\label{fig13}
\end{figure}

\begin{figure}[ht]
\begin{center}
\vspace{1cm}
\includegraphics[width=10.6cm]{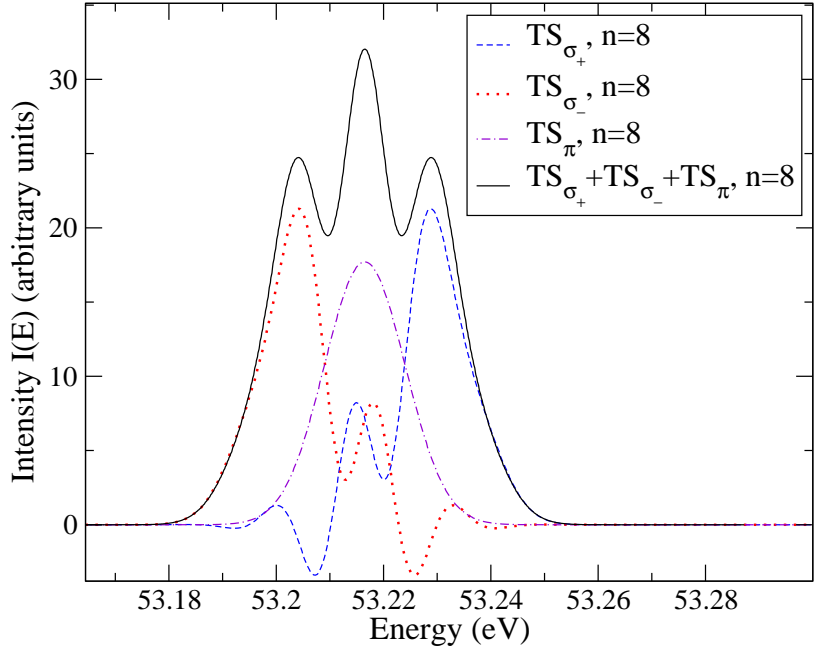}
\end{center}
\caption{(Color online) Modeling of a line $J=3\rightarrow\;J'=4$ of transition array Fe VII $3d^2\rightarrow 3d4p$ with Taylor-series expansion of eighth order, for each component separately for B=1.5 MG and $v$=5 10$^{-5}$ eV$^2$.}
\label{fig14}
\end{figure}

Throughout the paper, the calculations denoted ``exact'' are performed with the Flexible Atomic Code (FAC) code \cite{GU08}. Figure \ref{fig12} shows that, for $B$=1.25 MG and $v$=5 10$^{-5}$ eV$^2$, the TS expansion converges to the exact profile with a very good accuracy. Such an approach still works fairly well even when the profile starts to exhibit oscillations due to the important separation of the $\pi$, $\sigma_+$ and $\sigma_-$ components (see Fig. \ref{fig13}) for B=1.5 MG (corresponding to $\mu_BB/\sqrt{v}\approx1.23$). In the latter case however, the convergence is quite slow: a satisfactory agreement is still not achieved at the order $n$=16. The Taylor-series method is valid for $\mu_B B\lesssim\sqrt{v}$, but breaks down if $\mu_B B$ becomes much larger than $\sqrt{v}$. Note that expression (\ref{iq2}) can be exploited for a rough determination of the magnitude of the magnetic field $B$, provided that variance $v$ of the other broadening mechanisms is known (see appendix C). It is interesting to mention, as can be seen on Fig. \ref{fig14}, that the modeling of each component separately is not satisfactory at all, since in the present case, the separate TS expansions exhibit some oscillations and can even become negative, for $\sigma_+$ and $\sigma_-$ components. However, such variations do not affect the resulting total function (sum of the three components). 

\subsection{\label{subsec41} A-type Gram-Charlier expansion series}

An alternative to the Taylor-series method consists in using a statistical distinction based on the Gram-Charlier development. Once the centered moments $\mu_{n,c}$ of a discrete distribution $A(E)$ are known, such a distribution can be modeled using an analytical function which preserves an arbitrary number of these moments. It is possible to build a function using the properties of orthogonal polynomials and their associated basis functions \cite{KENDALL69,GILLERON08,PAIN09a,PAIN10a}. The A-type Gram-Charlier (GC) expansion series is a combination of products of Hermite polynomials by a Gaussian function:

\begin{equation}
GC_n(E)=\frac{\exp\left(-\frac{y^2}{2}\right)}{\sqrt{2\pi \mu_{2,c}[A]}}
\left(1+\sum_{k=2}^{n}c_{k}~\text{He}_{k}(y)\right),
\end{equation}

with

\begin{equation}\label{eq:ck}
c_k=\sum_{j=0}^{\text{int}(k/2)}\frac{(-1)^j}{j!(k-2j)!2^j}~\alpha_{k-2j}[A],
\end{equation}

where $y=(E-\mu_1)/\sqrt{\mu_{2,c}}$, $n$ is the number of moments, $\text{int}(k/2)$ is the integer part of $k/2$ and the Hermite polynomial He$_k$ is defined in the preceding subsection \ref{subsec42}. The GC series uses the reduced centered moments $\alpha_n[A]$ of $A(E)$, which are defined by:

\begin{equation}\label{alpn}
\alpha_n[A]=\frac{\mu_{n,c}[A]}{(\mu_{2,c}[A])^{n/2}}.
\end{equation}

The fourth-order GC series reads:

\begin{eqnarray}\label{gc4}
GC_4(E)&=&\frac{\exp\left(-\frac{u^2}{2}\right)}{\sqrt{2\pi\mu_{2,c}[A]}}\left\{1-\frac{\alpha_3}{2}\left(u-\frac{u^3}{3}\right)\right.\nonumber\\
& &\left.+\frac{(\alpha_4-3)}{24}(3-6 u^2+u^4)\right\}.
\end{eqnarray}

\begin{table}[t]
\begin{center}
\begin{tabular}{|c|c|c|c|}\hline\hline
Line & $\alpha_3$ & $\alpha_5$ & $\alpha_7$ \\ \hline\hline
$^5F_1\rightarrow~ ^5F_2$ & 0.994 & 5.521 & 27.913 \\ 
$^7D_3\rightarrow~ ^7D_4$ & 0.889 & 6.067 & 41.822 \\ 
$^4D_{3/2}\rightarrow~ ^4D_{5/2}$ & 0.939 & 5.856 & 35.177 \\ 
$^5P_2\rightarrow~ ^5P_3$ & 0.913 & 5.977 & 38.670 \\\hline\hline
\end{tabular}
\end{center}
\caption{Parameters of the $\sigma$ component for several lines of the transition array Fe VII $3d^2\rightarrow 3d4p$. Odd reduced centered moments.}\label{tab6}
\end{table}

The truncated series $GC_n(E)$ may be viewed as a Gaussian function multiplied by a polynomial which accounts for the effects of departure from normality. Therefore it may be a slowly converging series when $A(E)$ differs strongly from the Gaussian distribution. It is also known to suffer from numerical instability since Eq. (\ref{eq:ck}) involves a sum of large terms of alternating sign. Still assuming a Gaussian physical broadening (see Eq. (\ref{gausbroa})) of the lines, the moments of the convolution read:

\begin{eqnarray}
\mu_{n,c}[A\otimes\Psi]&=&\frac{1}{\sqrt{\pi}}\sum_{k=0}^n\bin{n}{k}(2v)^{\frac{n-k}{2}}\Gamma\left(\frac{n-k+1}{2}\right)\nonumber\\
& &\times\left(\frac{1+(-1)^{n-k}}{2}\right)\mu_{k,c}[A],
\end{eqnarray}

where $x\mapsto \Gamma(x)$ is the usual Gamma function.

\begin{table}[t]
\begin{center}
\begin{tabular}{|c|c|c|c|c|}\hline\hline
Line & $\mathcal{V}$ & $\alpha_4$ & $\alpha_6$ & $\alpha_8$ \\ \hline\hline
$^5F_1\rightarrow~ ^5F_2$ & 0.4500 & 2.778 & 12.654 & 62.592 \\ 
$^7D_3\rightarrow~ ^7D_4$ & 2.2500 & 3.032 & 16.426 & 114.19 \\ 
$^4D_{3/2}\rightarrow~ ^4D_{5/2}$ & 0.7875 & 2.914 & 14.637 & 87.850 \\ 
$^5P_2\rightarrow~ ^5P_3$ & 1.2000 & 2.976 & 15.575 & 101.273 \\\hline\hline
\end{tabular}
\end{center}
\caption{Parameters of the $\sigma$ component for several lines of the transition array Fe VII $3d^2\rightarrow 3d4p$. Even reduced centered moments.}\label{tab7}
\end{table}

\subsubsection{\label{subsubsec412} Global Gram-Charlier expansion series for the total intensity}

In that case, $A=I=\sum_{q=-1}^{+1}I_q$ and
 
\begin{eqnarray}\label{mun}
\mu_{n,c}[I]&=&\frac{1}{3}\sum_{i=0}^n\sum_{j=0}^{n-i}\bin{n}{i}\bin{i}{j}\left(\eline\right)^{n-i-j}\nonumber\\
& &\times(\mu_BB)^{i+j}\sum_{q=-1}^{+1}\left(\mathcal{M}_1^{[q]}\right)^j\left(\mathcal{M}_{i,c}^{[q]}\right)^j.
\end{eqnarray}

\begin{figure}[ht]
\begin{center}
\vspace{1cm}
\includegraphics[width=10.6cm]{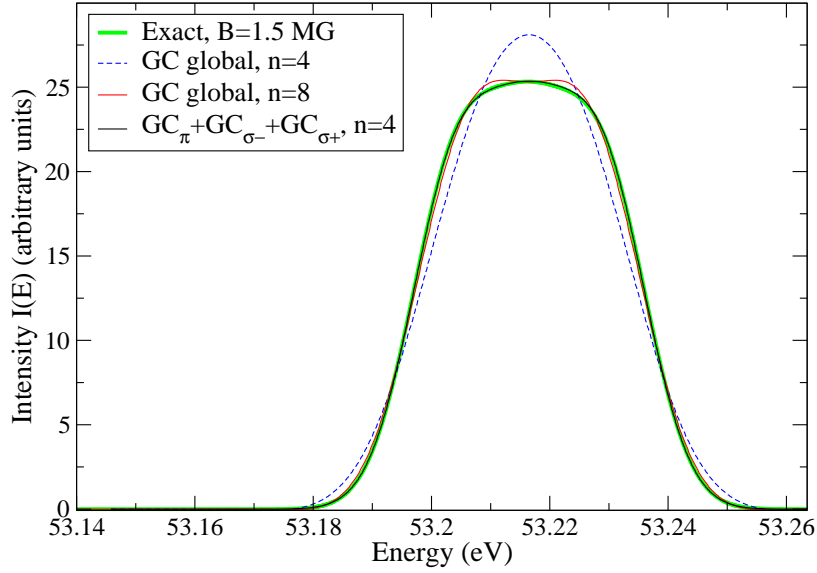}
\end{center}
\caption{(Color online) Modeling of a line $J=3\rightarrow\;J'=4$ of transition array Fe VII $3d^2\rightarrow 3d4p$ with A-type Gram-Charlier expansion series of different orders, compared to the exact calculation. $B$=1.5 MG and $v$=5 10$^{-5}$ eV$^2$. The sum of three fourth-order GC functions and the exact calculation are almost superimposed.}
\label{fig15}
\end{figure}

Figure \ref{fig15} shows that, for a line $J=3\rightarrow\;J'=4$ of transition array Fe VII $3d^2\rightarrow 3d4p$, the fourth-order A-type Gram-Charlier distribution $GC_4(E)$ of Eq. (\ref{gc4}) provides a satisfactory depiction of the profile. However, when the order increases, the departure from the exact calculation becomes larger and larger. This is due to the fact that the reduced centered moments $\alpha_n[I]$ (see Eq. (\ref{alpn})) do not depend on $B$. Therefore, such an approach can be applied only if the global shape $I(E)$ is close to a Gaussian, \emph{i.e.} does not have a non-monotonic character. This implies that the method is valid only if $\mu_BB<\sqrt{v}$, so that the $\pi$ and $\sigma_{\pm}$ components are not too separated. This approach provides a good depiction of the profile if $\mu_BB < \sqrt{v}$. 

\subsubsection{\label{subsubsec411} A-type Gram-Charlier expansion series for each component}

In that case, $A=I_q$ and

\begin{equation}
\mu_{n,c}[I_q]=\mathcal{M}_{n,c}^{[q]}\times(\mu_BB)^n.
\end{equation}

\begin{figure}[ht]
\begin{center}
\vspace{1cm}
\includegraphics[width=10.6cm]{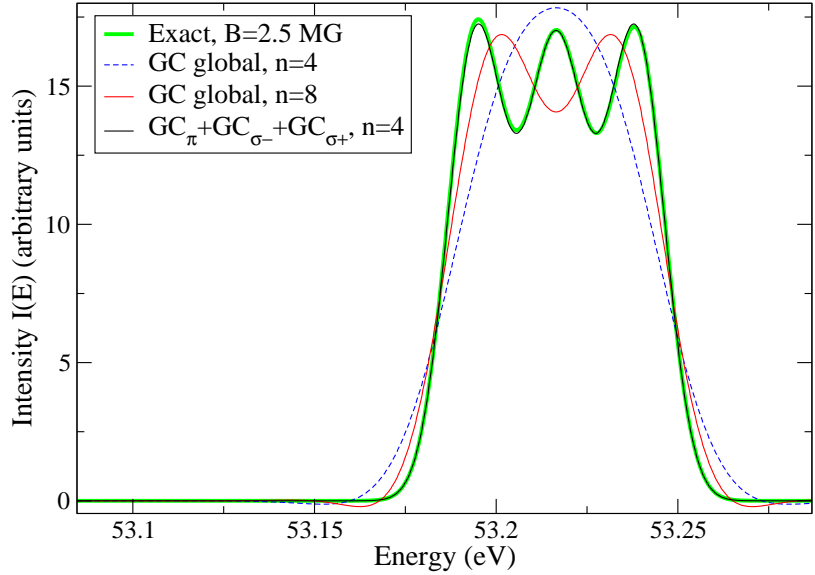}
\end{center}
\caption{(Color online) Modeling of a line $J=3\rightarrow\;J'=4$ of transition array Fe VII $3d^2\rightarrow 3d4p$ with A-type Gram-Charlier expansion series of different orders, compared to the calculation. $B$=2.5 MG and $v$=5 10$^{-5}$ eV$^2$. The sum of three fourth-order GC functions and the exact calculation are almost superimposed.}
\label{fig16}
\end{figure}

\begin{figure}[ht]
\begin{center}
\vspace{1cm}
\includegraphics[width=10.6cm]{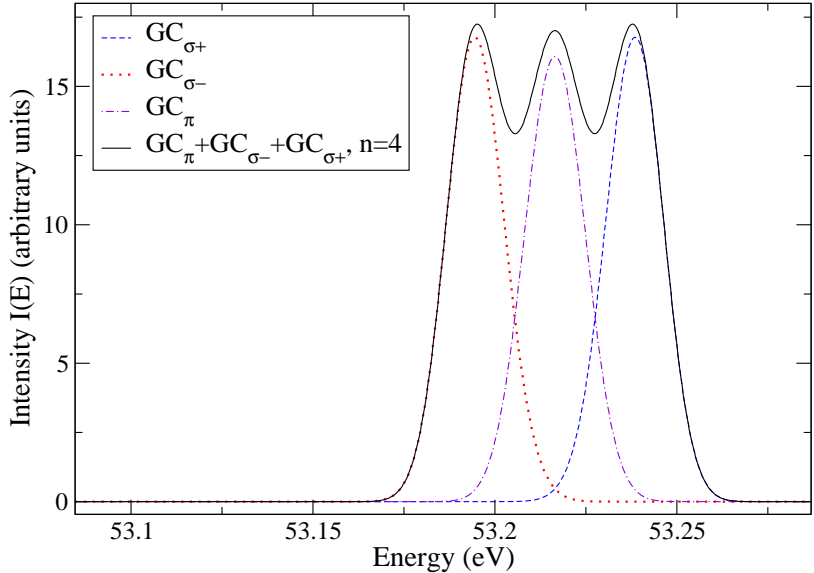}
\end{center}
\caption{(Color online) Modeling of a line $J=3\rightarrow\;J'=4$ of transition array Fe VII $3d^2\rightarrow 3d4p$ with A-type Gram-Charlier expansion series of fourth order for each component. $B$=2.5 MG and $v$=5 10$^{-5}$ eV$^2$.}
\label{fig17}
\end{figure}

This approach has a wider validity range than the previous one (see Fig. \ref{fig16} the case of a magnetic field equal to $B$=2.5 MG). When the ratio $\mu_BB/\sqrt{v}$ becomes larger than one, the summation of three A-type Gram-Charlier expansion series brings more flexibility. One can notice on the wings that A-type Gram-Charlier expansion series yield negative values in certain circumstances. However, it provides a good global depiction of the profile. Figure \ref{fig17} displays the modeling of each component separately. The $\sigma_+$ and $\sigma_-$ profiles do not show the oscillations observed with the TS expansion (see Fig. \ref{fig14}).

\section{\label{sec5} Global accounting for Zeeman effect on a transition array}

\subsection{Statistical description}

The absorption and emission spectra consist of a huge number of electric-dipolar (E1) lines. A transition array \cite{HARRISON31} represents all the E1 lines between two configurations and is characterized by a line-strength-weighted distribution of photon energy $E$:

\begin{eqnarray}\label{def:A(E)}
I(E)&=&\sum_{\gamma J\rightarrow\gamma'J',M,M',q}I_q(E)\nonumber\\
&=&\sum_{\gamma J\rightarrow\gamma'J'}~~S_{\gamma J,\gamma'J'}~\Psi(E-\eline).
\end{eqnarray}

The sum runs over the upper and lower levels of each line belonging to the transition array. 

In the UTA (Unresolved Transition Arrays) approach \cite{BAUCHE79}, the discrete set of lines (as $\delta$ functions) is replaced by a continuous function (usually Gaussian) which preserves its first- and second-order moments. The moments of this distribution are evaluated as

\begin{equation}
\mu_n\approx\frac{\sum_{\gamma J\rightarrow\gamma'J'}S_{\gamma J,\gamma'J'}~\eline^n}{\sum_{\gamma J\rightarrow\gamma'J'}S_{\gamma J,\gamma'J'}}.
\end{equation}

It is possible to derive analytical formulae for the moments $\mu_n$ using Racah's quantum-mechanical algebra and second-quantization techniques of Judd \cite{JUDD67}. Such expressions, which depend only on radial integrals, have been published by Bauche-Arnoult \emph{et al.} \cite{BAUCHE79,BAUCHE82,BAUCHE84,BAUCHE85} for the moments $\mu_{n,c}$ (centered moments with respect to $\mu_1$) with $n\leq 3$ of several kinds of transition arrays (relativistic or not). Karazija \emph{et al.} have proposed an algorithm in order to calculate the moments of a transition array using diagrammatic techniques \cite{KARAZIJA91a, KARAZIJA91b, KARAZIJA95}.

The contribution of Zeeman effect to the $k^{th}$-order moment of a transition array for a polarization $q$ reads \cite{DALLOT96}

\begin{eqnarray}
\mu^Z_{k}&=&\sum_{\gamma J M,\gamma'J'M'}\mathcal{P}_{\gamma J M\rightarrow \gamma'J'M'}\nonumber\\
& &\times\left\{\eline+\mu_BB\;(g_{\gamma'J'}M'-g_{\gamma J}M)\right\}^k,
\end{eqnarray}

$\mathcal{P}_{\gamma J M\rightarrow \gamma'J'M'}$ being the probability of a transition from $(J,M_J)$ to $(J',M_J-q)$ (component). Using the binomial development, one obtains:

\begin{equation}
\mu^Z_k=\mu_k+\sum_{p=1}^k(\mu_BB)^p\bin{k}{p}\mu_{k-p}\mathcal{F}_p,
\end{equation}

where 

\begin{equation}
\mu_k=\sum_{\gamma J M, \gamma' J' M'}\mathcal{P}_{\gamma J M\rightarrow \gamma'J'M'}(E_{\gamma'J'}-E_{\gamma J})^k
\end{equation}

and

\begin{equation}
\mathcal{F}_p=\sum_{\gamma J M, \gamma' J' M'}(g_{\gamma'J'}M'-g_{\gamma J}M)^p\;\mathcal{P}_{\gamma J M\rightarrow \gamma'J'M'},
\end{equation}

which can be evaluated using the techniques mentioned in the preceding sections \ref{sec3} and \ref{sec4}. 
 
The complexity of such a calculation encouraged us to develop an alternative approximate method. Suppose one wants to include the effet of a magnetic field in a numerical code devoted to the computation of opacity or emissivity, without performing the diagonalization of the Zeeman Hamiltonian. The numerical code can be either based on a detailed (see sections \ref{sec2}, \ref{sec3} and \ref{sec4}) or a statistical description (relying on the UTA formalism as mentioned above). The main contribution comes from the splitting of the line into three components. Indeed, if one considers 3 components with zero width positioned at $\eline-\mu_BB$, $\eline$ and $\eline+\mu_BB$ (each having the same strength $S_{\gamma J,\gamma'J'}$), the variance is equal to:

\begin{eqnarray}
& &\frac{1}{S_{\gamma J,\gamma'J'}}\left(\frac{S_{\gamma J,\gamma'J'}}{3}(\eline-\mu_BB-\eline)^2\right.\nonumber\\
& &+\frac{S_{\gamma J,\gamma'J'}}{3}(\eline-\eline)^2\nonumber\\
& &\left.+\frac{S_{\gamma J,\gamma'J'}}{3}(\eline+\mu_BB-\eline)^2\right)
\end{eqnarray}

which is equal to $2/3\;(\mu_BB)^2\approx 3.35\; 10^{-5}$ [$B$(MG)]$^2$. 

The broadening of each $q$ component separately due to the magnetic field (which is larger for a $\pi$ than for a $\sigma$ component as a consequence of Eqs. (\ref{ineg1}) and (\ref{ineg2})) is always much smaller than $2/3\;(\mu_BB)^2$ (by at least one order of magnitude). Thus, the contribution of a magnetic field to an UTA can be taken into account roughly by adding a contribution $2/3\;(\mu_BB)^2$ to the statistical variance. In case of a detailed transition array, the Zeeman broadening of a line can be represented by a fourth-order A-type Gram-Charlier expansion series (Eq. (\ref{gc4})), \emph{i.e.}: 

\begin{eqnarray}\label{newpro}
\Psi_Z(E-\eline)&=&\sum_{q=-1}^1\frac{\exp\left(-\frac{y_q^2}{2}\right)}{\mu_BB\sqrt{2\pi\mathcal{M}_{2,c}^{[q]}}}\nonumber\\
& &\times\left\{1-\frac{\alpha_3^{[q]}}{2}\left(y_q-\frac{y_q^3}{3}\right)\right.\nonumber\\
& &\left.+\frac{(\alpha_4^{[q]}-3)}{24}(3-6 y_q^2+y_q^4)\right\},\nonumber\\
\end{eqnarray}

where 

\begin{equation}
y_q=\frac{E-\eline-q~g_e~\mu_BB}{\mu_BB\sqrt{\mathcal{M}_{2,c}^{[q]}}}.
\end{equation}

The coefficient $g_e$ of the line $\gamma J\rightarrow\gamma'J'$ is given by
 
\begin{eqnarray}
g_e&=&\frac{1}{4}\;\left\{2(g_{\gamma J}+g_{\gamma'J'})\right.\nonumber\\
& &\left.+(g_{\gamma J}-g_{\gamma'J'})(J-J')(J+J'+1)\right\},
\end{eqnarray}

where $g_{\gamma J}$ and $g_{\gamma'J'}$ are the Land\'e factors of levels $\gamma J$ and $\gamma'J'$ respectively \cite{RACAH65,MARTIN78,BIEMONT10}. 

\subsection{Approximation of the coefficient $g_e$}

If the values of $g_{\gamma J}$ and $g_{\gamma'J'}$ are unknown, we suggest to replace $g_e$ by its average value in LS coupling $\bar{g}_e$. Knowing the distribution of spectroscopic terms $Q(S,L)$ \cite{BAUCHE87,GILLERON09}, it is possible to get a quick estimate of $\bar{g}_e$. Indeed, the equality 

\begin{equation}
\sum_{\gamma(SL)J}X_{SLJ}=\sum_{S,L}Q(S,L)X_{SLJ},
\end{equation}

where $X_{SLJ}$ is any quantity depending on $S$, $L$ and $J$, enables one to deal with the coupling of angular momenta $L$ and $S$ avoiding the use of coefficients of fractional parentage. One has

\begin{equation}\label{varta}
\bar{g}_e=\sum_{S,L,J}\;\sum_{L',J'}\;Q(S,L')\;g_e(S,L,J,L',J')\;\epsilon(L,L',J,J'),
\end{equation}

where $\epsilon(L,L',J,J')$ stands for the selection rules: $L'=L, L-1$ or $L+1$ avoiding $L'=L=0$ and $J'=J, J-1$ or $J+1$ avoiding $J'=J=0$. One has

\begin{eqnarray}
g_e(S,L,J,L',J')&=&\frac{1}{4}\;\left\{2(g_{SLJ}+g_{SL'J'})\right.\nonumber\\
& &+(g_{SLJ}-g_{SL'J'})\nonumber\\
& &\left.\times(J-J')(J+J'+1)\right\},
\end{eqnarray}

where the Land\'{e} factors are estimated in LS coupling:

\begin{eqnarray}
g_{SLJ}&=&1+\frac{(g_s-1)(\bar{J}+\bar{S}-\bar{L})}{2\bar{J}}\nonumber\\
& &=\frac{g_s+1}{2}+\frac{(g_s-1)(\bar{S}-\bar{L})}{2\bar{J}}\,
\end{eqnarray}

with the convention of Biedenharn \emph{et al.} \cite{BIEDENHARN52}, $\bar{x}=x(x+1)$. The quantity $g_s$ represents the anomalous gyromagnetic ratio defined in section \ref{sec1}. Assuming $g_s\approx 2$, one has

\begin{equation}\label{landels}
g_{SLJ}=\frac{3}{2}+\frac{(\bar{S}-\bar{L})}{2\bar{J}}.
\end{equation}

Table \ref{tab8} contains values of the Land\'e factor calculated in LS coupling using Eq. (\ref{landels}) as well as $g_e(S,L,J,L',J')$ factor for different lines.

\begin{table}[ht]
\begin{center}
\begin{tabular}{|c|c|c|c|}\hline\hline 
Line & $g_{SLJ}$ & $g_{SL'J'}$ & $g_e(S,L,J,L',J')$ \\\hline\hline
$^5F_1\rightarrow~ ^5F_2$ & 0 & 1 & 1.5 \\ 
$^7D_1\rightarrow~ ^7D_2$ & 3 & 2 & 1.5 \\ 
$^4D_{3/2}\rightarrow~ ^4D_{5/2}$ & 1.2 & 1.371 & 1.417 \\ 
$^5P_2\rightarrow~ ^5P_3$ & 1.833 & 1.667 & 1.5 \\\hline\hline
\end{tabular}
\end{center}
\caption{Land\'{e} factors for several lines evaluated from formula (\ref{landels}).}\label{tab8}
\end{table}

The problem of listing the terms arising in a complex configuration can be solved from elementary group theory \cite{BREIT26,CURL60,KARAYIANIS65,KATRIEL89,XU06}. The number $Q(S,L)$ of LS terms of a configuration $\ell^N$ can be obtained from the relation

\begin{equation}\label{recsl}
Q(S,L)=\sum_{M_S=S}^{S+1}\sum_{M_L=L}^{L+1}(-1)^{S-M_S+L-M_L}P_N(M_S,M_L),
\end{equation}

where $P_N(a,b)$, number of states with a given $M_S=a$ and $M_L=b$, can be obtained using recursive formulas \cite{GILLERON09}:

\begin{eqnarray}\label{slrec}
& &P_N(M_S,M_L)=\frac{1}{N}\sum_{i=1}^G\sum_{k=1}^N(-1)^{k+1}\nonumber\\
& &\times P_{N-k}\left(M_S-k\frac{(-1)^i}{2};M_L-km_i+\frac{(-1)^i}{2}\right),\nonumber\\
& &
\end{eqnarray}

where $G=\sum_{i=1}^NG_i$, $G_i$ being the degeneracy of orbital $i$. For the non-relativistic configuration $\ell^N$:

\begin{equation}
m_k=\frac{2k-4\ell-3+(-1)^k}{4}\;\;;\;\;1\leq k\leq 4\ell+2
\end{equation}

and for the relativistic configuration $j^N$:

\begin{equation}
m_k=k-j+1\;\;;\;\;1\leq k\leq 2j+1.
\end{equation}

The recurrence (\ref{slrec}) is initialized with

\begin{equation}
P_0(M_S,M_L)=\delta(M_S)\delta(M_L).
\end{equation}

For a configuration $\ell_1^{N_1}\ell_2^{N_2}\ell_3^{N_3}\cdots$, $P(M_S,M_L)$ is determined through the relation

\begin{equation}
P_{N_1,N_2,\cdots}(M_S,M_L)=(P_{N_1}\otimes P_{N_2}\otimes \cdots)(M_S,M_L),
\end{equation}

where the distributions are convolved two at a time, which means

\begin{eqnarray}
(P_{N_i}\otimes P_{N_j})(M_S,M_L)=\;\;\;\;\;\;\;\;\;\;\;\;\;\;\;\;\;\;\;\;\;\;\;\;\;\;\;\;\;\;\;\;\;& &\nonumber\\
\sum_{M_S'=-\infty}^{+\infty}\sum_{M_L'=-\infty}^{+\infty}P_{N_i}(M_S',M_L')\;\;\;\;\;\;\;\;\;\;\;\;\;\;\;\;& &\nonumber\\
\times P_{N_j}(M_S-M_S',M_L-M_L').& &
\end{eqnarray}

\begin{figure}[ht]
\begin{center}
\vspace{1cm}
\includegraphics[width=10.6cm]{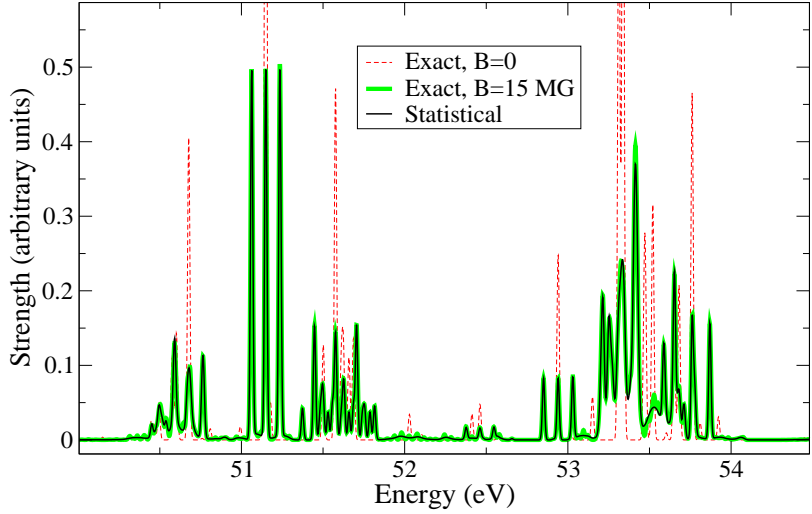}
\end{center}
\caption{(Color online) Effect of a 15 MG magnetic field on transition array Fe VII $3d^2\rightarrow 3d4p$ with a convolution width of 0.017 eV. The curves corresponding to the exact ($B$=15 MG)  and statistical calculation are almost superimposed.}
\label{fig18}
\end{figure}

\begin{figure}[ht]
\begin{center}
\vspace{1cm}
\includegraphics[width=10.6cm]{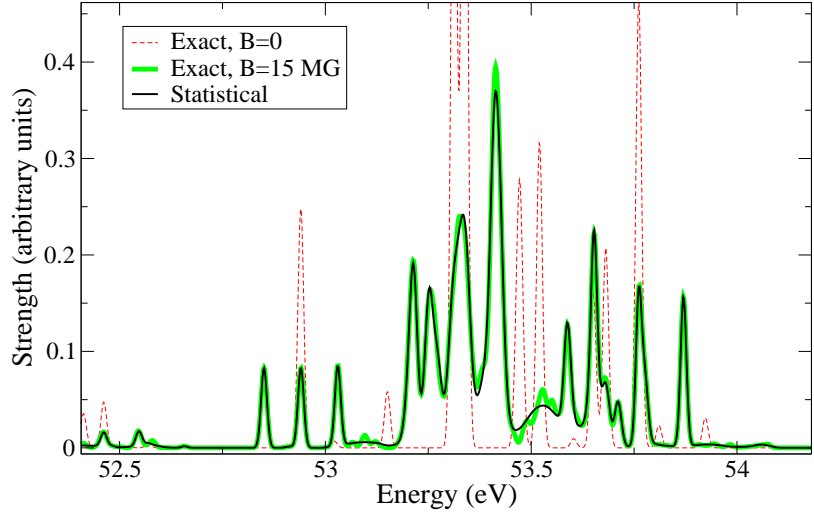}
\end{center}
\caption{(Color online) Detail of Fig. \ref{fig18}.}
\label{fig19}
\end{figure}

\begin{figure}[ht]
\begin{center}
\vspace{1cm}
\includegraphics[width=10.6cm]{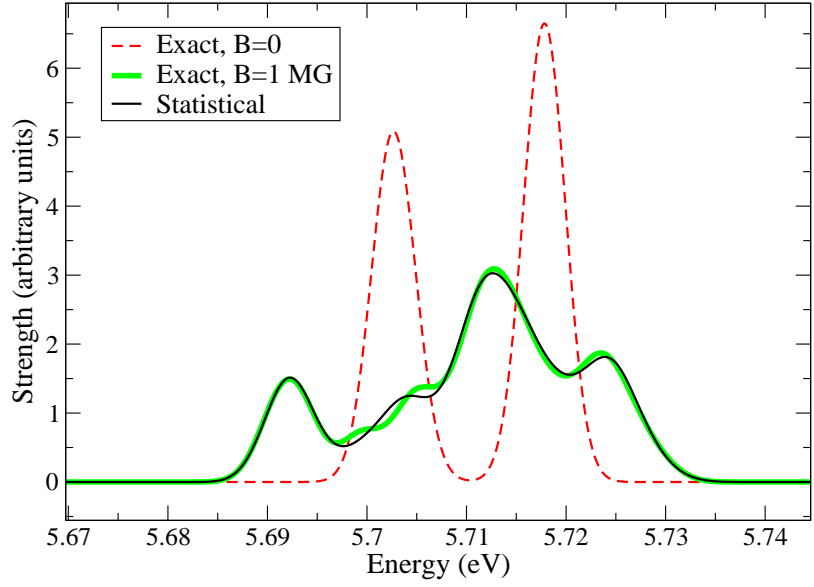}
\end{center}
\caption{(Color online) Effect of a 1 MG magnetic field on C V $1s2s~^3S\rightarrow 1s2p~^3P$ triplet transition with a convolution width of 0.005 eV.}
\label{fig20}
\end{figure}

Thus, in order to take into account approximately the impact of the magnetic field when the number of lines is large, we suggest to convolve the transition array in the absence of a magnetic field with the distribution of Eq. (\ref{newpro}). Figures \ref{fig18}, \ref{fig19} and \ref{fig20} (example taken from Mc Lean \cite{MCLEAN84, GRIEM97}) show that the results are quite close to the exact calculation. The main approximation here comes from the fact that $g_e$ is replaced by its average value in LS coupling, which is justified in case of very strong magnetic fields (Paschen-Back effect). Table \ref{tab9} displays the energies and Land\'e factors of the levels of configurations $1s2s$ and $1s2p$ in intermediate coupling and table \ref{tab10} indicates the oscillator strength multiplied by the degeneracy $g_{\gamma J}f_{\gamma J,\gamma' J'}$ of the six lines.

\begin{table}[ht]
\begin{center}
\begin{tabular}{|c|c|c|c|c|}\hline\hline
Level number & $J$ & Energy (eV) & Land\'e (IC) & Configuration \\\hline\hline
1 & 0 & 303.99067 & 1.500000000 & $1s2s$ \\       
2 & 1 & 297.88824 & 2.002320051 & $1s2s$ \\       
3 & 0 & 303.59204 & 1.500000000 & $1s2p$ \\       
4 & 1 & 303.59057 & 1.501152782 & $1s2p$ \\        
5 & 1 & 307.61242 & 1.000007244 & $1s2p$ \\
6 & 2 & 303.60607 & 1.501160026 & $1s2p$ \\\hline\hline                
\end{tabular}
\end{center}
\caption{Energy (relative to the energy of $1s^2$) and Land\'e factor of the different levels of configurations $1s2s$ and $1s2p$.}\label{tab9}
\end{table}

\begin{table}[ht]
\begin{center}
\begin{tabular}{|c|c|c|}\hline\hline
Initial level & Final level & $g_{\gamma J}f_{\gamma J,\gamma' J'}$ \\\hline\hline
4 & 1 & 1.56936 10$^{-7}$ \\
1 & 5 & 9.47301 10$^{-2}$ \\
2 & 3 & 3.99278 10$^{-2}$ \\
2 & 4 & 0.11969 \\
2 & 5 & 3.80306 10$^{-6}$ \\
2 & 6 & 0.20016 \\\hline\hline
\end{tabular}
\end{center}
\caption{E1 lines in the transition array $1s2s\rightarrow 1s2p$ in intermediate coupling.}\label{tab10}
\end{table}   

If needed, the evaluation of $\bar{g}_e$ can be refined. For instance, it is possible to calculate an average value of $g_{\gamma J}$ depending only on $J$. This can be achieved using the sum rule \cite{COWAN81}

\begin{equation}
\sum_{\gamma}g_{\gamma J}=\sum_{\alpha LS}g_{LSJ}\sum_{\gamma J}\langle\gamma J|\alpha LSJ\rangle^2=\sum_{\alpha LS}g_{LSJ},
\end{equation}

which states that the sum of the Land\'e factors for any given $J$ is independent of the coupling conditions. Such a property stems from the fact that the trace of a matrix is invariant under an orthogonal transformation. One can thus define an average Land\'e factor associated to a given value of $J$:

\begin{equation}
\bar{g}_J=\frac{\sum_{\gamma}g_{\gamma J}}{Q(J)}=\frac{\sum_{\alpha LS}g_{LSJ}}{Q(J)},
\end{equation}

where $Q(J)$ is the number of levels having angular momentum $J$ \cite{CONDON35,BAUCHE87}, which can be evaluated recursively \cite{GILLERON09}, in a similar manner to $Q(S,L)$ (see Eq. (\ref{recsl})).

\section{\label{sec6} Hyperfine structure}

The same methodology can be applied in order to determine analytically the moments of the hyperfine components of a line. The hyperfine operator in the subspace corresponding to the relevant nucleus and atomic level reads:

\begin{equation}
H_m=A_J(\vec{I}.\vec{J}),
\end{equation}

where $A_J$ is the magnetic hyperfine-structure constant of the level $\gamma J$. The $n^{th}$-order moment of the hyperfine components is provided by the expression

\begin{eqnarray}
\mathcal{M}_n&=&\frac{1}{S_{FF'MM'}}\sum_{F,F',M,M'}\left[\langle\gamma JIFM|H_m|\gamma JIFM\rangle\right.\nonumber\\
& &\left.-\langle\gamma' J'IF'M'|H_m|\gamma' J'IF'M'\rangle\right]^n\nonumber\\
& &\times\langle\gamma JIFM|\mathcal{Z}_q^{(1)}|\gamma JIFM\rangle^2,
\end{eqnarray}

where $\mathcal{Z}_q^{(1)}$ is the $q$-component of the dipole operator $\mathcal{Z}^{(1)}$. The $J$-file sum rule \cite{CONDON35} enables one to simplify the expression of the strength:

\begin{eqnarray}
S_{FF'MM'}&=&\sum_{F,F',M,M'}\langle\gamma JIFM|\mathcal{Z}_q^{(1)}|\gamma JIFM\rangle^2\nonumber\\
&=&\sum_{F,F',M,M'}\troisjm{F}{1}{F'}{-M}{q}{M'}^2\langle F||\mathcal{Z}^{(1)}||F'\rangle^2\nonumber\\
&=&\frac{1}{3}\sum_{F,F'}\langle F||\mathcal{Z}^{(1)}||F\rangle^2\nonumber\\
&=&\frac{1}{3}\sum_{F}[F]=\frac{1}{3}[I,J],
\end{eqnarray}

and therefore

\begin{eqnarray}\label{mn0}
\mathcal{M}_n&=&\frac{1}{2^n[I,J]}\sum_{F,F'}\left(A_JX_{FIJ}-A_{J'}X_{F'IJ'}\right)^n\nonumber\\
& &\times\langle F||\mathcal{Z}^{(1)}||F'\rangle^2,
\end{eqnarray}

where $X_{FIJ}=(-1)^F~(\bar{F}-\bar{I}-\bar{J})$. Equation (\ref{mn0}) can be written

\begin{eqnarray}
\mathcal{M}_n&=&\frac{1}{2^n[I,J]}\sum_{F,F'}[F,F']\left(A_JX_{FIJ}-A_{J'}X_{F'IJ'}\right)^n\nonumber\\
& &\times\langle (IJ)F||\mathcal{Z}^{(1)}||(IJ)F'\rangle^2,
\end{eqnarray}

or

\begin{eqnarray}
\mathcal{M}_n&=&\frac{1}{2^n[I]}\sum_{F,F'}[F,F']\left(A_JX_{FIJ}-A_{J'}X_{F'IJ'}\right)^n\nonumber\\
& &\times\sixj{F}{1}{F'}{J'}{I}{J}^2.
\end{eqnarray}

In the case where $F$ or $F'$ is equal to 0, the calculation is very simple \cite{HUHNERMANN93}. In the general case, using

\begin{equation}
X_{FIJ}=2(-1)^{F+I+J}\sqrt{\bar{I}\bar{J}[I,J]}\sixj{F}{J}{I}{1}{I}{J},
\end{equation}

one has to calculate:

\begin{eqnarray}
& &\sum_{F,F'}(-1)^{k_1F+k_2F'}[F,F']\sixj{F}{J}{I}{1}{I}{J}^{k_1}\nonumber\\
& &\times\sixj{F'}{J'}{I}{1}{I}{J'}^{k_2}\sixj{F}{1}{F'}{J'}{I}{J}^2,
\end{eqnarray}

which can be done using graphical methods \cite{VARSHALOVICH88}. Another approach consists in adopting another point of view, leading to the evaluation of quantities of the type:

\begin{equation}
S_n=\sum_F[F](\bar{F}-a)^n
\end{equation}

where $a$ is a constant (depending on other quantum numbers). Such a quantity can be expressed, as for the Zeeman effect, in terms of Bernoulli numbers (see appendix B):

\begin{eqnarray}
S_n&=&\sum_{k=0}^n\sum_{j=0}^k\bin{n}{k}\bin{k}{j}a^{n-k}\nonumber\\
& &\times\left\{(2(B_{k+j+2}(I+J+1)-B_{k+j+2})\right.\nonumber\\
& &+B_{k+j+1}(I+J+1)-B_{k+j+1})/(I+J+1)\nonumber\\
& &+(2(B_{k+j+2}(|I-J|+1)-B_{k+j+2})\nonumber\\
& &\left.+B_{k+j+1}(|I-J|+1)-B_{k+j+1})/(|I-J|+1)\right\}.\nonumber\\
\end{eqnarray}

The splitting of $F$ components in a weak magnetic field \cite{SOBELMAN72} is in every way similar to the splitting of $J$ levels. The scale of the splitting is determined by the factor $g_F$, which is defined by

\begin{equation}
\langle\gamma JIFM|H_z|\gamma JIFM|\rangle=\mu_BBg_{\gamma J}\frac{\bar{F}+\bar{J}-\bar{I}}{2\bar{F}}M,
\end{equation}

and connected with the Land\'e factor by

\begin{equation}
g_F=g_{\gamma J}\frac{\bar{F}+\bar{J}-\bar{I}}{2\bar{F}}.
\end{equation}

\section{\label{sec7} Conclusion}

In this work, a statistical modeling of electric dipolar lines in the presence of an intense magnetic field was proposed. The formalism requires the moments of the Zeeman components of a line $\gamma J\rightarrow \gamma'J'$, which can be obtained analytically in terms of the quantum numbers and Land\'{e} factors. It was found that the fourth-order A-type Gram-Charlier expansion series provides better results than the usual development in powers of the magnetic field often used in radiative-transfer models. Using our recently published recursive method for the numbering of LS-terms of an arbitrary configuration, a simple approach to estimate the contribution of a magnetic field to the width (and higher-order moments) of a transition array of E1 lines was presented. We hope that such results will be useful for the interpretation of Z-pinch absorption or emission spectra, for the study of laser-induced magnetic fields in inertial-fusion studies, for the modeling of magnetized stars as well as for any application involving magnetic fields in spectroscopic studies of atomic and molecular systems. 

\vspace{0.5cm}

\normalsize {\bf Acknowledgments}

\vspace{0.5cm}

The authors would like to thank C. Bauche-Arnoult, J. Bauche and R. Karazija for helpful discussions.

\section{Appendix A: Expressions involving three-$jm$ and six-$j$ symbols used in sections \ref{sec3} and \ref{sec4}}

\begin{equation}
[a, b, c, \cdots]=(2a+1)(2b+1)(2c+1)\cdots
\end{equation}

\begin{equation}
\left\{\begin{array}{l}
\bar{x}=x(x+1)\\
h=\bar{J}-\bar{J'}'+2\\
\end{array}\right.
\end{equation}

\begin{eqnarray}\label{rs1}
&\sum_{M,M'}(-1)^{J-M}&\nonumber\\
&\times\troisjm{J}{1}{J'}{-M}{-q}{M'}\troisjm{J'}{1}{J}{-M'}{q}{M}\troisjm{J}{1}{J}{-M}{0}{M}&\nonumber\\
&=(-1)^{J+J'-q}\troisjm{1}{1}{1}{-q}{0}{q}\sixj{J'}{J}{1}{1}{1}{J}.&
\end{eqnarray}

\begin{eqnarray}\label{rs2}
\sum_{M,M'}\troisjm{J}{1}{J}{-M}{0}{M}\troisjm{J}{1}{J'}{-M}{-q}{M'}& &\nonumber\\
\times\troisjm{J'}{1}{J}{-M'}{q}{M}\troisjm{J}{1}{J}{-M}{0}{M}& &\nonumber\\
=\sum_{J''}(-1)^{J+J'+1+J''}[J'']\troisj{1}{J''}{1}\troisjm{1}{J''}{1}{-q}{0}{q}& &\nonumber\\
\times\sixj{J}{J}{1}{J''}{1}{J}\sixj{J'}{J}{1}{J''}{1}{J}.\nonumber\\& &
\end{eqnarray}

\begin{equation}\label{tj1}
\troisjm{1}{1}{1}{-q}{0}{q}=(-1)^{q+1}\frac{q}{\sqrt{6}}.
\end{equation}

\begin{equation}\label{tj2}
\troisjm{1}{J''}{1}{0}{0}{0}=\frac{(-1)^{J''}(J''-1)(J''+2)}{\sqrt{(2-J'')!(J''+3)!}}.
\end{equation}

\begin{equation}\label{sj1}
\sixj{J'}{J}{1}{0}{1}{J}=\frac{(-1)^{J+J'+1}}{\sqrt{3[J]}}.
\end{equation}

\begin{equation}\label{sj2}
\sixj{J'}{J}{1}{1}{1}{J}=(-1)^{J+J'}\frac{h}{2\sqrt{6[J]\bar{J}}}.
\end{equation}

\begin{equation}\label{sj3}
\sixj{J'}{J}{1}{2}{1}{J}=(-1)^{J+J'+1}\frac{3\;h(h-1)-8\bar{J}}{\sqrt{120[J](2J-1)(2J+3)\bar{J}}}.
\end{equation}

\begin{equation}\label{tj5}
\troisjm{J}{1}{J}{-M}{0}{M}=(-1)^{J-M}\frac{M}{\sqrt{[J]\bar{J}}}.
\end{equation}

\section{Appendix B: Bernoulli polynomials and numbers}

The Bernoulli polynomials can be obtained by successive derivation of a generating function:

\begin{equation}
B_n(x)=\left.\frac{\partial^n}{\partial t^n}\left(\frac{t~\exp\left(xt\right)}{\exp(t)-1}\right)\right|_{t=0}.
\end{equation}

One can write

\begin{equation}
B_n(x)=\sum_{k=0}^n\bin{n}{k}B_k(0)\;x^{n-k},
\end{equation}

where $B_n(0)$ is the $n$-order Bernoulli number, which is non-zero only if $n$ is even and which can be obtained from the relation:

\begin{equation}
B_n(0)=-\frac{1}{n+1}\sum_{k=0}^{n-1}\bin{n}{k}B_k(0).
\end{equation}

The first Bernoulli polynomials are

\begin{equation}
B_0(x)=1, 
\end{equation}

\begin{equation}
B_1(x)=x-\frac{1}{2},
\end{equation}

\begin{equation}
B_2(x)=x^2-x+\frac{1}{6},
\end{equation}

\begin{equation}
B_3(x)=(x-1)(x-\frac{1}{2})x,
\end{equation}

and

\begin{equation}
B_4(x)=x^4-2x^3+x^2-\frac{1}{30}.
\end{equation}

The Bernoulli polynomials obey the following identity:

\begin{equation}\label{beid}
\sum_{k=1}^nk^p=\frac{B_{p+1}(n+1)-B_{p+1}(0)}{p+1},
\end{equation}

and the Bernoulli numbers have the explicit Laplace's determinantal formula \cite{KORN67}:

\begin{equation}
B_n(0)=\left| \begin{array}{ccccc}
1 & 0 & \cdots & 0 & 1 \\
\frac{1}{2!} & 1 &  & 0 & 0 \\
\vdots &  & \ddots &  & \vdots \\
\frac{1}{n!} & \frac{1}{(n-1)!} &  & 1 & 0 \\
\frac{1}{(n+1)!} & \frac{1}{n!} & \cdots & \frac{1}{2!} & 0 \\
\end{array}\right|.
\end{equation}

\section{Appendix C: diagnostic of the magnetic field}

Using second-order TS expansion (\ref{iq2}) and assuming the knowledge of the variance $v$ of the other broadening mechanisms, it becomes possible to estimate the magnitude of the magnetic field from the measurement of the full width at half maximum (FWHM) of the line $\delta=\text{FWHM}/(2\sqrt{v})$. 

\begin{equation}\label{est}
B=\frac{1}{\mu_B}\sqrt{\frac{v(1-2e^{-\delta^2/2})}{C(\theta)\left(1-2e^{-\delta^2/2}(1-\delta^2)\right)}},
\end{equation}

where

\begin{equation}
C(\theta)=A(\theta)~\left(\mathcal{M}_1^{[\sigma_+]}\right)^2+\mathcal{M}_{2,c}^{[\sigma_+]}+D(\theta)\mathcal{M}_{2,c}^{[\pi]},
\end{equation}

with

\begin{equation}
A(\theta)=\left(\frac{1+\cos^2(\theta)}{4}\right)
\end{equation}

and

\begin{equation}
D(\theta)=\frac{\sin^2(\theta)}{4}.
\end{equation}

This simple formula (\ref{est}) can provide an estimation of the magnetic field, even if the other broadening mechanisms (Stark, electron collisions, Doppler, autoionization) are dominant. However, it is not as efficient as the method proposed by Stambulchik \emph{et al.} \cite{STAMBULCHIK07}, which is applicable in situations where the magnetic field has various directions and amplitudes (or if they vary in time).


\begin{thebibliography}{99}

\bibitem{BABCOCK60} H. W. Babcock, Astrophys. J. \textbf{521}, 132 (1960).

\bibitem{BLACKETT47} P. M. S. Blackett, Nature \textbf{159}, 658 (1947).

\bibitem{ZEEMAN1896a} P. Zeeman, Versl. K. Akad. Wet. Amsterdam \textbf{5}, 181 (1896).

\bibitem{ZEEMAN1897} P. Zeeman, Ap. J. \textbf{5}, 332 (1897).

\bibitem{RYABCHICOVA06} T. Ryabchikova, O. Kochukhov, D. Kudryavtsev, I. Romanyuk, E. Semenko, S. Bagnulo, G. Lo Curto, P. North and M. Sachkov, Astron. Astrophys. \textbf{445}, L47 (2006).

\bibitem{LINDMAN10} E. L. Lindman, High Energy Density Phys. \textbf{6}, 227 (2010).

\bibitem{KIRKPATRICK95} R. C. Kirkpatrick, I. R. Lindemuth and M. S. Ward, Fusion Technol. \textbf{27}, 201 (1995).

\bibitem{STAMBULCHIK10} E. Stambulchik and Y. Maron, High Energy Density Phys. {\bf 6}, 9 (2010).

\bibitem{FERRI11} S. Ferri, A. Calisti, C. Moss{\'e}, L. Mouret, B. Talin, M. A. Gigosos, M. A. Gonz\'alez and V. Lisitsa, Phys. Rev. E {\bf 84}, 026407 (2011).

\bibitem{CALISTI10} A. Calisti, C. Moss{\'e}, S. Ferri, B. Talin, F. Rosmej, L. A. Bureyeva and V. S. Lisitsa, Phys. Rev. E {\bf 81}, 016406 (2010).

\bibitem{GILLERON09} F. Gilleron and J.-C. Pain, High Energy Density Phys. \textbf{5}, 320 (2009).

\bibitem{LANDE21} A. Land\'e, Z. Physik \textbf{5}, 231 (1921).

\bibitem{PASCHEN21} F. Paschen and E. Back, Physica \textbf{1}, 261 (1921).

\bibitem{GODBERT09} L. Godbert-Mouret, J. Rosato, H. Capes, Y. Marandet, S. Ferri, M. Koubiti, R. Stamm, M. Gonzales and M. Gigosos, High Energy Density Phys. \textbf{5}, 162 (2009).

\bibitem{NGUYENHOE67} Nguyen-hoe, H.-W. Drawin and L. Herman, J. Quant. Spectrosc. Radiat. Transfer \textbf{7}, 429 (1967).

\bibitem{KENDALL69} M. G. Kendall and A. Stuart, {\it Advanced Theory of Statistics} (Hafner, New York, 1969), Vol. 1.

\bibitem{VARSHALOVICH88} D. A. Varshalovich, A. N. Moskalev and V. K. Khersonskii, {\it Quantum Theory of Angular Momentum} (World Scientific, Singapore, 1988).

\bibitem{JUCYS62} A. Jucys, Y. Levinson and V. Vanagas, {\it Mathematical Apparatus of the Theory of Angular Momentum} (Israel Program for Scientific Translations, Jerusalem, 1962).

\bibitem{ELBAZ69} E. El-Baz, J. Lafoucri\`{e}re and B. Castel, {\it Traitement graphique de l'alg\`{e}bre des moments angulaires}, Collection de monographies de physique (Masson et Cie, Paris, 1969), in french.

\bibitem{ELBAZ72} E. El-Baz and B. Castel, {\it Graphical Methods of Spin Algebras} (Marcel Dekker, New-York, 1972).

\bibitem{ELBAZ85} E. El-Baz, {\it Alg\`{e}bre de Racah et analyse vectorielle graphiques} (Ellipses, Marketing Editions, 1985), in french.

\bibitem{BIEDENHARN52} L. C. Biedenharn, J. M. Blatt and M. E. Rose, Rev. Mod. Phys. {\bf 24}, 249 (1952).

\bibitem{SHENSTONE29} A. G. Shenstone and H. A. Blair, Phil. Mag. \textbf{8}, 765 (1929).

\bibitem{EDMONDS57} A. R. Edmonds, {\it Angular Momentum in Quantum Mechanics} (Princeton Univ. Press, Princeton, 1957).

\bibitem{LANDI85} E. Landi Degl'Innocenti, Solar Phys. \textbf{99}, 1 (1985).

\bibitem{BAUCHE88b} J. Bauche and J. Oreg, J. Physique Colloque C1 \textbf{49}, 263 (1988).

\bibitem{KARAZIJA91a} R. Karazija, {\it Sums of Atomic Quantities and Mean Characteristics of Spectra} (Mokslas, Vilnius, 1991), in russian.

\bibitem{KARAZIJA91b} R. Karazija, Acta Phys. Hungarica \textbf{70}, 367 (1991).

\bibitem{KARAZIJA95} R. Karazija and S. Kucas, Lith. J. Phys. \textbf{35}, 155 (1995).

\bibitem{BORDARIER70} Y. Bordarier, {\it Contribution \`a l'emploi de m\'ethodes graphiques en spectroscopie atomique}, University of Paris, Orsay (1970), in french.

\bibitem{BAR88} A. Bar-Shalom and M. Klapisch, Comput. Phys. Comm. \textbf{50}, 375 (1988).

\bibitem{OREG90} J. Oreg, W. H. Goldstein, A. Bar-Shalom and M. Klapisch, J. Comp. Phys. \textbf{91}, 460 (1990).

\bibitem{MATHYS87a} G. Mathys and J. O. Stenflo, Astron. Astrophys. \textbf{171}, 368 (1987).

\bibitem{MATHYS87b} G. Mathys and J. O. Stenflo, Astron. Astrophys. Suppl. Ser. \textbf{67}, 557 (1987).

\bibitem{GU08} M. F. Gu, Can. J. Phys. \textbf{86}, 675 (2008).

\bibitem{GILLERON08} F. Gilleron, J.-Ch. Pain, J. Bauche and C. Bauche-Arnoult, Phys. Rev. E \textbf{77}, 026708 (2008).

\bibitem{PAIN09a} J.-Ch. Pain, F. Gilleron, J. Bauche and C. Bauche-Arnoult, High Energy Density Phys. \textbf{5}, 294 (2009).

\bibitem{PAIN10a} J.-Ch. Pain, F. Gilleron, J. Bauche and C. Bauche-Arnoult, High Energy Density Phys. \textbf{6}, 356 (2010).

\bibitem{HARRISON31} G. R. Harrison and M.H. Johnson, Phys. Rev. \textbf{38}, 757 (1931).

\bibitem{BAUCHE79} C. Bauche-Arnoult, J. Bauche and M. Klapisch, Phys. Rev. A \textbf{20}, 2424 (1979).

\bibitem{JUDD67} B. R. Judd, {\it Second Quantization and Atomic Spectroscopy} (Johns Hopkins University, Baltimore, 1967).

\bibitem{BAUCHE82} C. Bauche-Arnoult, J. Bauche and M. Klapisch, Phys. Rev. A \textbf{25}, 2641 (1982).

\bibitem{BAUCHE84} C. Bauche-Arnoult, J. Bauche and M. Klapisch, Phys. Rev. A \textbf{30}, 3026 (1984).

\bibitem{BAUCHE85} C. Bauche-Arnoult, J. Bauche and M. Klapisch, Phys. Rev. A \textbf{31}, 2248 (1985).

\bibitem{DALLOT96} P. Dallot, Phys. Rev. A \textbf{53}, 4007 (1996).

\bibitem{RACAH65} N. Zeldes, Arch. Hist. Exact Sci. \textbf{63}, 289 (2009). G. Racah developed a method for fitting by least squares to the theoretical formulas the measured g-factors of atomic levels together with the energy values (see [N. Zeldes, Arch. Hist. Exact Sci. \textbf{63}, 289 (2009)] and references therein). Unfortunately, the work was never presented (G. Racah died in August 1965, before the conference).

\bibitem{MARTIN78} W. C. Martin, R. Zalubas and L. Hagan, {\it Atomic Energy Levels - The Rare Earth Elements}, NSRDS-NBS (U. S. Govt. Printing Off., Washington, D. C., 1978).

\bibitem{BIEMONT10} \'E. Bi{\'e}mont, P. Palmeri and P. Quinet, J. Phys. B: At. Mol. Opt. Phys. \textbf{43}, 074010 (2010).

\bibitem{BAUCHE87} J. Bauche and C. Bauche-Arnoult, J. Phys. B: At. Mol. Phys. \textbf{20}, 1659 (1987).

\bibitem{BREIT26} G. Breit, Phys. Rev. \textbf{28}, 334 (1926).

\bibitem{CURL60} R. F. Curl Jr. and J. E. Kilpatrick, Am. J. Phys. \textbf{28}, 357 (1960).

\bibitem{KARAYIANIS65} N. Karayianis, J. Math. Phys. \textbf{6}, 1204 (1965).

\bibitem{KATRIEL89} J. Katriel and A. Novoselsky, J. Phys. A: Math. Gen. \textbf{22}, 1245 (1989).

\bibitem{XU06} R. Xu and Z. Dai, J. Phys. B: At. Mol. Opt. Phys. \textbf{39}, 3221 (2006).

\bibitem{MCLEAN84} E. A. McLean, J. A. Stamper, C. K. Manka, H. R. Griem, D. W. Droemer and B. H. Ripin, Phys. Fluids {\bf 27}, 1327 (1984).

\bibitem{GRIEM97} H. R. Griem, {\it Principles of Plasma Spectroscopy} (Cambridge University Press, Cambridge, 1997).

\bibitem{COWAN81} R. D. Cowan, {\it The Theory of Atomic Structure and Spectra} (University of California Press, Berkeley, 1981).

\bibitem{CONDON35} E. U. Condon and G. H. Shortley, {\it The theory of atomic spectra} (Cambridge University Press, Cambridge, 1935).

\bibitem{HUHNERMANN93} H. H\"uhnermann, Phys. Scr. \textbf{T47}, 70 (1993).

\bibitem{SOBELMAN72} I. I. Sobelman, {\it Introduction to the theory of atomic spectra} (Pergamon, New York, 1972).

\bibitem{KORN67} G. A. Korn and T. M. Korn, {\it Mathematical Handbook for Scientists and Engineers}, Section 21.5, (Mac Graw Hill, 1967).

\bibitem{STAMBULCHIK07} E. Stambulchik, K. Tsigutkin and Y. Maron, Phys. Rev. Lett. {\bf 98}, 225001 (2007).

\end{thebibliography}
\end{document}